\documentclass[a4paper,12pt]{article}
\usepackage{amssymb}
\usepackage{amsfonts}
\usepackage{amsmath}
\usepackage{color}
\usepackage[left=3.4cm, right=3.4cm, top=3.2cm, bottom=3.2cm]{geometry}

\newtheorem{theorem}{Theorem}

\newtheorem{corollary}[theorem]{Corollary}

\newtheorem{lemma}[theorem]{Lemma}

\newtheorem{proposition}[theorem]{Proposition}

\input{tcilatex}

\begin{document}

\title{Approximate Optimality and the Risk/Reward Tradeoff in a Class of
Bandit Problems\thanks{%
Chen is at School of Mathematics, Shandong University, zjchen@sdu.edu.cn,
Epstein is at Department of Economics, McGill University,
larry.epstein@mcgill.ca, and Zhang is at School of Mathematics, Shandong
University, zhang$\_$gd@mail.sdu.edu.cn. Chen gratefully acknowledges the
support of the National Key R\&D Program of China (grant No. ZR2019ZD41),
and the Taishan Scholars Project. Zhang gratefully acknowledges the support
of Shandong Provincial Natural Science Foundation, China (grant No.
ZR2022QA063 and ZR2021MA098).}}
\author{Zengjing Chen \and Larry G. Epstein \and Guodong Zhang}
\maketitle
\date{}

\begin{abstract}
This paper studies a sequential decision problem where payoff distributions
are known and where the riskiness of payoffs matters. Equivalently, it
studies sequential choice from a repeated set of independent lotteries. The
decision-maker is assumed to pursue strategies that are approximately
optimal for large horizons. By exploiting the tractability afforded by
asymptotics, conditions are derived characterizing when specialization in
one action or lottery throughout is asymptotically optimal and when
optimality requires intertemporal diversification. The key is the constancy
or variability of risk attitude, that is, the decision-maker's risk/reward
tradeoff.

\bigskip

\end{abstract}

\section{Introduction}

We study the following sequential choice problem. There are $K$ arms (or
actions), each yielding a random payoff. Payoff distributions are
independent across arms and identical and independent for a given arm across
distinct trials. At each stage $i=1,2,...,n$, the decision-maker (DM) must
choose one arm, knowing both the realized payoffs from previous choices and
the distribution of the payoff for each arm. She chooses a strategy ex ante
specifying future contingent choices. This is a special case of a bandit
problem, whence the usage of `arm' rather than `action.' Alternatively, the
decision problem can be viewed as choice of a dynamic strategy when facing a
repeated set of (single-stage) gambles or lotteries, where each lottery is
repeated independently. Thus it is an instance of dynamic risk management.

Because we are interested in varying horizons, we define a strategy for an
infinite horizon, and then use its truncation for any given finite horizon.%
%
%
%
%
%
%
%
%
%
\ Refer to a strategy as \emph{asymptotically optimal }if the expected
utility it implies in the limit as horizon $n\rightarrow \infty $ is at
least as large as that implied by any other strategy; or equivalently, if it
is \emph{approximately optimal for large horizons}. We study large-horizon
approximations to the value (indirect utility) of the sequential choice
problem and also corresponding asymptotically optimal strategies. Our focus
is on the derivation of analytical (as opposed to computational) results,
particularly with regard to the effect of risk non-neutrality. For example,
we demonstrate that (non)constancy of risk attitude, suitably measured,
determines whether specialization in a single arm throughout or
diversification across time is asymptotically optimal. 

Consider three concrete settings that fit our model well. \textit{Gambling}:
A gambler chooses sequentially which of several given slot machines to play. 
\textit{News site}: Each visitor to a site decides whether to click
depending on the news header presented to her. The website (DM) chooses the
header (arm) with clicks being the payoffs. Users are drawn independently
from a fixed distribution. \noindent \textit{Ad selection}: A website (DM)
displays an ad (arm) for each visitor, who is an i.i.d. draw as above. If
she clicks, the payoff to the website is a predetermined price, depending on
the ad and paid by the advertiser. Importantly for the fit with our model,
in all three settings payoffs are realized quickly after an arm is chosen,
and plausibly a large number of trials occur in a relatively short period of
time.\footnote{%
Daily life provides other repeated choice problems, for example, which
transportation mode or route to use to get to work, though the longer time
interval between choices suggests a poorer fit with the model.}

We have two related reasons for studying asymptotics. First, from the
modeler's perspective, it promotes tractability and the derivation of
analytical results. Bandit problems are notoriously difficult to solve
analytically, as opposed to numerically, in the presence of nonindifference
to risk. A second reason for studying asymptotics is that tractability may
be a concern also for the decision-maker within the model who faces an
extremely complicated large-horizon optimization problem. In such
circumstances, she may seek a strategy that is approximately optimal if her
horizon is sufficiently long. The presumption that a large-horizon heuristic
can alleviate cognitive limitations is supported by two features of our
results: (i) asymptotic optimality depends on payoff distributions and the
values they induce \emph{only through their means and variances} (Theorem %
\ref{thm-bandits}), that is, \emph{DM need not know more about the
distributions}; and (ii) also by the relative simplicity of the explicit
asymptotically optimal strategies in some cases (Theorem \ref{thm-tradeoff}).

The focus on asymptotics leads to other noteworthy features of our analysis.
First, unsurprisingly, it leads to our exploiting limit theorems, most
notably a central limit theorem (CLT). The classical CLT considers a
sequence $\left( X_{i}\right) $ of identically and independently distributed
random variables, hence having a fixed mean and variance, which assumptions
are adequate for evaluation of the repeated play of a single arm, and hence
also for addressing the once-and-for-all choice between arms. However, in
the more economically relevant case of sequential choice, we must evaluate
strategies which permit switching between arms, and hence also between
payoff distributions, at any stage. Accordingly, in our key technical
result, CLT (Proposition \ref{propn-CLT}), means and variances of $\left(
X_{i}\right) $ can vary with $i$ subject only to the restriction that they
lie in a fixed set.

The role played by limit theorems is reflected also in our specification of
the utility index $u$. We adopt a form of multiattribute utility theory
(Keeney and Raiffa 1993), whereby two attributes of random payoff streams
are assumed to be important. Accordingly, $u:\mathbb{R}^{2}\longrightarrow 
\mathbb{R}$ has two arguments, namely the sample average and the $\sqrt{n}$%
-weighted average of deviations from conditional means, exactly the
statistics whose limiting distributions are the focus in the LLN (law of
large numbers) and CLT respectively. The function $u$ itself is restricted
only by technical conditions. Nevertheless, the resulting model is both
tractable and also flexible enough to accommodate interesting special cases.
As an example of the diversity of cases accommodated, one is a form of
mean-variance for our sequential setting, and another essentially replaces
variance by semivariance. The differing implications of these two
specifications illustrate one message that the paper is intended to convey:
the mean-variance model exhibits constant risk attitude and accordingly
predicts specialization in one arm, that is, time-diversification is not
important in sufficiently large horizons, while risk attitude varies
endogenously in the mean-semivariance model which therefore predicts
specialization only for some but not all parameter values.

The paper proceeds as follows. Related literature is discussed next. The
model and main results follow in Section \ref{section-model}. Most proofs
are provided in the Appendix, which also contains our CLT. Proofs of some
details are collected in the Online Appendix.

\subsection{Related literature}

Decision-making in the presence of repeated gambles has been studied in
several papers. We mention some that help to locate this paper in the
context of this literature. In McCardle and Winkler (1992), a coin with
uncertain bias is tossed repeatedly. The decision-maker observes the
outcomes of all tosses, updates her beliefs accordingly, and chooses
sequentially how much to bet on heads at each history.%
%
Full Bayesian rationality is assumed. The authors argue that some of the
model's predictions about willingness to bet are unintuitive and they
attribute this to the assumption that future betting opportunities are fully
anticipated and incorporated via optimization. Accordingly, they suggest the
need for simplifying heuristics that still accommodate some, but not all,
"grand world" considerations, though no specific heuristics are proposed. We
share the broad view that dynamic decision problems under uncertainty are
exceedingly complex and propose, for our setting, the simplification
consisting of approximate optimality for large horizons. In Gollier (1996),
a single lottery (with known distribution) is repeated independently, and
the decision-maker accepts or rejects the lottery at each stage. Choice is
determined by maximization of the expected utility of terminal wealth. His
paper and ours address different questions. Gollier focuses on how the
option to gamble in the future affects the willingness to gamble today,
while we are focussed on behavior in the remote future because it describes
approximately optimal behavior for sufficiently long horizons. Another
difference is that his setting with a riskless option can be viewed as the
special case of our setting where there are two lotteries at each stage and
where one is degenerate (attaches probability 1 to the outcome $0$). The
assumption that there is only one risky asset (or lottery) and one riskless
is common in finance. However, it is restrictive and it is not clear if and
how Gollier's analysis would extend. Both options being risky poses
significant technical complications for the modeler and cognitive challenges
for the decision-maker within the model.

Samuelson (1963) identified as fallacious the reliance on the law of large
numbers as justifying acceptance of any sufficiently long sequence of
repetitions of a positive mean bet even if the single bet is rejected, and
suggested that it indicated undue attention to the variance associated with
multiple repetitions. While we do not address Samuelson's fallacy here, we
see in it the hint that there could be a role for the other major limit
theorem, the CLT, in the broader study of risk-taking given repeated
gambles. In that sense, this paper is inspired by Samuelson (1963). Related
is the literature examining the effect on financial risk-taking of horizon
length (e.g. of age in a life-cycle portfolio context), for example, whether
a longer horizon promotes risk-taking because it offers a greater
possibility to smooth out risks over time (see, for example, Samuelson
(1989) and Gollier and Zeckhauser (2002)). We differ from this literature in
(at least) two respects. First, we model behavior in the long-horizon limit;
we do not study the effect of differing horizon length on risk-taking. A
second critical difference is that while in the finance literature, assets
are divisible and can be combined into portfolios at any stage, the
lotteries available to our decision-maker are indivisible and only one can
be chosen at any stage. Consequently, portfolio diversification is excluded
herein while diversification over time is feasible and a focus.\footnote{%
When both kinds of diversification are feasible, Samuelson (1989,1997)
argues that time-diversification is inferior. Here we explore whether
time-diversification is useful for long horizon planning in settings where
portfolio diversification is not feasible.}

Approximate optimality for long horizons has been studied in finance in the
context of portfolio turnpike theorems (see, for example, Huberman and Ross
(1983) and the references therein). This literature studies the conditions
under which wealth-independent (hence "constant") portfolios are
approximately optimal for sufficiently long horizons. Accordingly, an
important factor is the relation between wealth and the benefit from
diversifying across assets at any given time. In contrast, in our model at
any instant the decision-maker can choose a single lottery from the given
finite set, and thus only diversification over time is feasible. In
addition, we study approximate optimality without imposing any form of
constancy; for example, Theorem \ref{thm-strategies}(v) illustrates the case
of an asymptotically optimal strategy that is not constant across time (that
is, the gamble chosen varies with time).

All of the papers cited above assume maximization of the expected utility of
terminal wealth. As outlined earlier, we model payoffs and utility
differently. There are precedents for "nonstandard" utility specifications
in the context of repeated gambles; for example, alternatives to expected
utility theory are either adopted or advocated by Chew and Epstein (1988),
Benartzi and Thaler (1999) and Lopes (1996).

The other major connection is to the bandit literature since our decision
problem is the special case of a multi-armed bandit problem where payoff
distributions are known and hence need not be learned. Most of the
literature (see Berry and Fristedt (1985) and Slivkins (2022) for
textbook-like treatments) assumes a finite horizon and that choices are
driven by expected total rewards, that is, risk neutrality. Studies that
explicitly address risk attitudes include Sani, Lazaric and Munos (2013),
Zimin, Ibsen-Jensen and Chatterjee (2014), Vakili and Zhao (2016), and
Cassel, Manor and Zeevi (2021). They assume regret minimization rather than
expected utility maximization, and focus on computational algorithms rather
than on qualitative theoretical results. Further, they are motivated by the
nature of learning about unknown payoff distributions, and thus by the
exploration/exploitation tradeoff, while we assume known distributions and
focus instead on the risk/reward tradeoff. Though it is important to
understand both tradeoffs and their interactions, as an initial step we
focus on only one in this paper, that being the tradeoff for which there
exists very limited theoretical analysis. Theorem \ref{thm-tradeoff} gives
analytical results on the latter tradeoff by exploiting the advantages of
large-horizon approximations.

In a more technical vein, our CLT connects this paper to the literature on
nonlinear CLTs, that is CLTs where the expectations operator is nonlinear,
for example, because of the multiplicity of priors and where expectation is
defined by the infimum (or supremum) of expectations as one varies over all
priors. The infimum is typically motivated, as in the maxmin model (Gilboa
and Schmeidler 1989), by robustness to ambiguity or model uncertainty. The
nonlinear CLTs in Peng (2007, 2019) and Fang et al (2019) are motivated in
this way (see Peng (2019, Thm 2.4.8), for example). They do not make a
connection to Bayesian sequential decision-making, nor is such a connection
apparent in their work. In contrast, the decision-maker in our model is
Bayesian and does not perceive ambiguity. Nevertheless, a set of probability
measures arises (implicitly) from the multiplicity of arms and strategies,
and a supremum applies because of utility maximization over the set of
strategies, or equivalently, over the probability measures they induce.
Chen, Epstein and Zhang (2023) introduced the use of a nonlinear CLT to
model Bayesian decision-makers. It differs from the present paper both
technically and in its economic focus as explained following the statement
of our CLT (Proposition \ref{propn-CLT}).

\section{The Model\label{section-model}}

\subsection{Preliminaries}

Let $(\Omega ,\mathcal{F},P)$ be the probability space on which all
subsequent random variables are defined. The random variables $X_{k}$, $%
1\leq k\leq K$, represent the random rewards from the $K$ arms, and $%
\{X_{k,n}:n\geq 1\}$ denote their independent and identically distributed
copies. We assume that each $X_{k}$ has a finite mean and variance, denoted
by 
\begin{equation}
\mu _{k}:=E_{P}[X_{k}],\ \sigma _{k}^{2}:=Var_{P}\left[ X_{k}\right] ,\quad
1\leq k\leq K.  \label{musigma}
\end{equation}%
The largest and smallest means and variances are given by%
\begin{align}
& \overline{\mu }=\max \{\mu _{1},\cdots ,\mu _{K}\},\ \ \underline{\mu }%
=\min \{\mu _{1},\cdots ,\mu _{K}\},  \label{bars} \\
& \overline{\sigma }^{2}=\max \{\sigma _{1}^{2},\cdots ,\sigma _{K}^{2}\},\ 
\underline{\sigma }^{2}=\min \{\sigma _{1}^{2},\cdots ,\sigma _{K}^{2}\}. 
\notag
\end{align}%
The set of mean-variance pairs is 
\begin{equation}
\mathcal{A}=\{\left( \mu _{k},\sigma _{k}^{2}\right) :1\leq k\leq K\}\text{.}
\label{A}
\end{equation}%
The convex hull of $\mathcal{A}$ is a convex polygon. Denote by $\mathcal{A}%
^{ext}$ its set of extreme points.

A \emph{strategy} $\theta $ is a sequence of $\{1,\cdots ,K\}$-valued random
variables, $\theta =(\theta _{1},\cdots ,\theta _{n},\cdots )$. $\theta $
selects arm $k$ at round $n$ in states for which $\theta _{n}=k$. Thus the
corresponding reward is $Z_{n}^{\theta }$ given by 
\begin{equation}
Z_{n}^{\theta }=X_{k,n}\text{ where }\theta _{n}=k\text{.}  \label{zntheta}
\end{equation}%
The strategy $\theta $ is \emph{admissible} if $\theta _{n}$ is $\mathcal{H}%
_{n-1}^{\theta }$-measurable for all $n\geq 1$, where%
\begin{equation*}
\mathcal{H}_{n-1}^{\theta }=\sigma \{Z_{1}^{\theta },\cdots ,Z_{n-1}^{\theta
},\theta _{1},...,\theta _{n-1}\}\text{ for }n>1\text{, and }\mathcal{H}%
_{0}^{\theta }=\{\emptyset ,\Omega \}\text{.}
\end{equation*}%
The information at stage $n$ captured by $\mathcal{H}_{n-1}^{\theta }$
includes both past choices of arms and the corresponding history of payoffs.
Allowing the arm chosen at stage $n$ to depend on past choices permits
strategies that alternate stochastically between arms. Given the serial
independence of payoffs, there is no learning rationale for conditioning on
past payoffs. However, \textit{past payoffs matter in general at any stage
because they may influence the attitude towards the risk associated with
current and future choices. }

The set of all admissible strategies is $\Theta $. (All strategies
considered below will be admissible, even where not specified explicitly.)%
\newline

\subsection{Utility}

For each horizon $n$, we specify the expected utility function $U_{n}$ used
to evaluate strategies $\theta $ and the payoff streams that they generate.
Let $u:\mathbb{R}^{2}\longrightarrow \mathbb{R}$ be the corresponding
von-Neumann Morgenstern (vNM) utility index and define $U_{n}$ by%
\begin{equation}
U_{n}\left( \theta \right) =E_{P}\left[ u\left( \frac{1}{n}%
\sum\limits_{i=1}^{n}Z_{i}^{\theta },\left( \sum\limits_{i=1}^{n}\frac{1}{%
\sqrt{n}}\left( Z_{i}^{\theta }-E_{P}[Z_{i}^{\theta }|\mathcal{H}%
_{i-1}^{\theta }]\right) \newline
\right) \right) \right] \text{.}  \label{Un}
\end{equation}%
The two arguments of $u$ correspond to the two attributes or characteristics
of a random payoff stream that DM takes into account. The first argument of $%
u$ is the sample average outcome under strategy $\theta $, and the second,
the $\sqrt{n}$-weighted average of deviations from conditional means,
represents sample volatility. Observe that the second argument has zero
expected value relative to the measure $P$. Though one might have expected
the term (as volatility) to be replaced by its square or by its absolute
value, the important point is that its evaluation be nonlinear, and here
nonlinearity enters via $u$. 
The presence of conditional rather than unconditional means reflects the
sequential nature of the setting. With regard to the $\sqrt{n}$-weighting,
as is familiar from discussions of the classical LLN and CLT, the scaling by 
$\frac{1}{n}$ implies that in large samples "too little" weight is given to
volatility (e.g. variance) relative to mean. Roughly, as described further
at the end of this section, the above specification \textit{models a
decision-maker who takes into account both mean and variance even
asymptotically}.

\medskip

\noindent \textbf{Remark}: As is familiar, a Savage act (random variable)
defined over a state space that is endowed with a probability measure
induces a lottery over outcomes. Similarly here, any strategy $\theta $
induces, via $P$, a multistage lottery, from which it follows that $\theta $
can be viewed as describing the sequential (or contingent) choice from a set
of repeated lotteries.

\medskip

Admittedly, the specification (\ref{Un}) is ad hoc in the sense of
(currently) lacking axiomatic foundations. We propose it because it seems
plausible and it delivers novel results. In addition, we are not aware of
any other model of preference over random payoff streams of arbitrary finite
length that has axiomatic foundations and that has something interesting to
say in our context. The special case of (\ref{Un}) where $u$\ is additively
separable and linear in its second argument (example (u.1) below) can be
axiomatized, but imposes a priori that only means matter asymptotically when
choosing between arms and hence is too special (Theorem \ref{thm-tradeoff}%
(v)). Take the further special case where $u$\ is also linear in its first
argument but where payoffs are denominated in utils. This is the expected
additive utility model (discounting can be added) that is the workhorse
model in economics. However, it does not work well in our setting, for
example, in the applied contexts in the introduction. We take the underlying
payoffs or rewards at each stage to be objective quantities, such as the
number of clicks or dollars. In all these cases, the relevant payoff when
choosing a strategy is the sum of single stage payoffs, e.g. the total
number of clicks, or in more formal terms, stage payoffs are perfect
substitutes. However, discounted expected utility with nonlinear stage
utility index models them as imperfect substitutes.

Utility has a particularly transparent form when $\theta =\theta ^{\mu
,\sigma }$ specifies choosing an arm described by the pair $\left( \mu
,\sigma ^{2}\right) $ repeatedly regardless of previous outcomes. In this
case payoffs are i.i.d. with mean $\mu $ and variance $\sigma ^{2}$. Thus
the conditional expectation appearing in (\ref{Un}) equals $\mu $, and the
classical LLN and CLT imply that in the large horizon limit risk is
described by the normal distribution $\mathbb{N}\left( 0,\sigma ^{2}\right) $
and 
\begin{equation}
\lim_{n\rightarrow \infty }U_{n}\left( \theta ^{\mu ,\sigma }\right) =\int
u\left( \mu ,\cdot \right) d\mathbb{N}\left( 0,\sigma ^{2}\right) \text{.}
\label{Uiid}
\end{equation}%
Consequently, if $u\left( \mu ,\cdot \right) $ is concave, then (asymptotic)
risk aversion is indicated in the sense that 
\begin{equation*}
\lim_{n\rightarrow \infty }U_{n}\left( \theta ^{\mu ,\sigma }\right) \leq
u\left( \mu ,0\right) \text{.}
\end{equation*}

Here are examples of utility indices $u$ and the implied utility functions $%
U_{n}$ that will be referred to again in the sequel.

\medskip

\noindent \textbf{Example (utility indices)}

\noindent \textbf{(u.1)} $u\left( x,y\right) =\varphi \left( x\right)
+\alpha y$. Then%
\begin{equation*}
U_{n}\left( \theta \right) =E_{P}\left[ \varphi \left( \frac{1}{n}%
\sum\limits_{i=1}^{n}Z_{i}^{\theta }\newline
\right) \newline
\right] \newline
\end{equation*}%
\textbf{(u.2)} $u\left( x,y\right) =\varphi \left( \left( 1-\alpha \right)
x+\alpha y\right) $, where $0<\alpha \leq 1$. Then 
\begin{equation*}
U_{n}\left( \theta \right) =E_{P}\left[ \varphi \left( \left( 1-\alpha
\right) \frac{1}{n}\sum\limits_{i=1}^{n}Z_{i}^{\theta }\newline
+\alpha \frac{1}{\sqrt{n}}\sum\limits_{i=1}^{n}\left( Z_{i}^{\theta
}-E_{P}[Z_{i}^{\theta }|\mathcal{H}_{i-1}^{\theta }]\right) \right) \newline
\right]
\end{equation*}%
\newline
\textbf{(u.3)} (Mean-variance) $u\left( x,y\right) =$\noindent\ $x-\alpha
y^{2}$, where $\alpha >0$. Then%
\begin{eqnarray}
U_{n}\left( \theta \right) &=&\frac{1}{n}E_{P}\left[ \sum%
\limits_{i=1}^{n}Z_{i}^{\theta }\right] -\alpha \frac{1}{n}Var_{P}\left[
\sum\limits_{i=1}^{n}\left( Z_{i}^{\theta }-E_{P}[Z_{i}^{\theta }|\mathcal{H}%
_{i-1}^{\theta }]\right) \newline
\right]  \label{Umv} \\
&=&\frac{1}{n}\sum\limits_{i=1}^{n}\left( E_{P}\left[ Z_{i}^{\theta }\right]
-\alpha Var_{P}\left[ Z_{i}^{\theta }-E_{P}[Z_{i}^{\theta }|\mathcal{H}%
_{i-1}^{\theta }]\right] \right) \text{,}  \notag
\end{eqnarray}%
\newline
which is a form of the classic mean-variance specification for our setting.%
\footnote{%
The second equality follows from the fact that, for $i\not=j$, $%
Z_{i}^{\theta }-E_{P}[Z_{i}^{\theta }|\mathcal{H}_{i-1}^{\theta }]$ and
\par
$Z_{j}^{\theta }-E_{P}[Z_{j}^{\theta }|\mathcal{H}_{j-1}^{\theta }]$ have
zero covariance under $P$.$\newline
$} For any arm with mean-variance pair $\left( \mu ,\sigma ^{2}\right) $
that is played repeatedly, 
\begin{equation}
U_{n}\left( \theta ^{\mu ,\sigma }\right) =\mu -\alpha \sigma ^{2}\text{,
for every }n\text{.}  \label{Un-iidquadratic}
\end{equation}%
\textbf{(u.4)} (Mean-semivariance) $u\left( x,y\right) =\noindent \ x-\alpha
y^{2}I_{(-\infty ,0)}\left( y\right) $. Only negative cumulative deviations
from (conditional) means are penalized. Then, given $\theta $ and letting $%
Y=\sum\limits_{i=1}^{n}\left( Z_{i}^{\theta }-E_{P}[Z_{i}^{\theta }|\mathcal{%
H}_{i-1}^{\theta }]\right) $, $Var_{P}\left[ Y\right] $ in (\ref{Umv}) is
replaced by the \emph{semivariance} $E_{P}\left[ Y^{2}I_{Y<0}\right] $. If $%
\theta =\theta ^{\mu ,\sigma }$, then 
\begin{equation*}
U_{n}\left( \theta ^{\mu ,\sigma }\right) \underset{n\rightarrow \infty }{%
\longrightarrow }\mu -\alpha \int_{-\infty }^{0}y^{2}d\mathbb{N}\left(
0,\sigma ^{2}\right) =\mu -\alpha \sigma ^{2}/2\text{.}
\end{equation*}

\noindent \textbf{(u.5)\ }(Shortfall penalty) $u\left( x,y\right) =\noindent
\ x-\alpha I_{(-\infty ,0)}\left( y\right) $. Only the existence of a
shortfall, and not its size, matters. Then 
\begin{eqnarray}
U_{n}\left( \theta ^{\mu ,\sigma }\right) &=&\mu -\alpha P\left( \frac{1}{%
\sqrt{n}}\sum\limits_{i=1}^{n}\left( Z_{i}^{\theta ^{\mu ,\sigma
}}-E_{P}[Z_{i}^{\theta ^{\mu ,\sigma }}|\mathcal{H}_{i-1}^{\theta ^{\mu
,\sigma }}]\right) \newline
<0\right)  \label{u5limit} \\
&&\underset{n\rightarrow \infty }{\longrightarrow }\mu -\alpha \mathbb{N}%
_{\left( 0,\sigma ^{2}\right) }(-\infty ,0)=\mu -\alpha /2\text{.}  \notag
\end{eqnarray}%
In particular, in the large horizon limit the utility of playing the single
arm $\left( \mu ,\sigma ^{2}\right) $ repeatedly does not depend on the
variance.

\medskip

\noindent \textbf{Remark}: For the last 3 examples, horizon length drops out
in the sense that maximizing $U_{n}\left( \theta \right) $ is equivalent to
maximizing the modified objective function $U_{n}^{\prime }\left( \theta
\right) $ defined as in (\ref{Un}) except that both $\frac{1}{n}$ and $\frac{%
1}{\sqrt{n}}$ are deleted. 

\medskip

Our model of utility provides a (local) measure of risk aversion, or
alternatively, of the mean-variance tradeoff, assuming that $u$ is suitably
differentiable (thus excluding examples (u.4) and (u.5)). {\tiny \ }Though
it is a slight variant of the well-known Arrow-Pratt measure (Pratt, 1964),
it might be worthwhile to derive it in our context. Consider a horizon equal
to $n$ stages and consider the choice for the last stage contingent on the
history represented by $\left( x,y\right) $, (partial sums corresponding to
the two averages in (\ref{Un})). Accordingly, DM uses the utility index $%
u\left( x+\cdot ,y+\cdot \right) $ to evaluate the next step. Consider her
evaluation of using the arm $\left( \epsilon ^{2}\mu ,\epsilon ^{2}\sigma
^{2}\right) $ for the final stage, where $\epsilon >0$ has the effect, when
small, of scaling down both the mean and variance of payoffs by $\epsilon
^{2}$. Using a second-order Taylor series approximation of $u\left( x+\cdot
,y+\cdot \right) $ about $\epsilon =0$, one obtains the expected utility%
\begin{equation*}
u\left( x,y\right) +\partial _{x}u\left( x,y\right) \frac{\epsilon ^{2}\mu }{%
n}+\frac{1}{2}\partial _{yy}^{2}u\left( x,y\right) \frac{\epsilon ^{2}\sigma
^{2}}{n}\text{.}
\end{equation*}%
Therefore, if we let%
\begin{equation}
\mu =\frac{-\frac{1}{2}\partial _{yy}^{2}u\left( x,y\right) }{\partial
_{x}u\left( x,y\right) }\sigma ^{2}\text{,}  \label{muAA}
\end{equation}%
then we can interpret $-\partial _{yy}^{2}u\left( x,y\right) /\partial
_{x}u\left( x,y\right) $ as approximating \textit{twice the mean-variance
ratio consistent with indifference to a small increase in risk}.

Two special cases are revealing. The measure of risk aversion is constant
for the mean-variance model:%
\begin{equation*}
\frac{-\frac{1}{2}\partial _{yy}^{2}u\left( x,y\right) }{\partial
_{x}u\left( x,y\right) }=\alpha \text{ \ for all }\left( x,y\right) \text{. }
\end{equation*}%
(See Theorem \ref{thm-tradeoff} and the ensuing discussion for behavioral
implications of this constancy.) Second, it is identically equal to $0$ for
(u.1), indicating risk neutrality in the sense defined by the measure, and
this is so regardless of the curvature of $\varphi $. More generally, the
measure does not involve $\partial _{xx}^{2}u\left( x,y\right) $, contrary
to what might be expected based on the Arrow-Pratt measure in expected
utility theory. As an "explanation" for this possibly puzzling feature, we
point out that $\partial _{xx}^{2}u\left( x,y\right) $ would appear in a
2nd-order Taylor series expansion if the added mean-variance pair (or arm)
were $\left( \epsilon \mu ,\epsilon ^{2}\sigma ^{2}\right) $ instead of $%
\left( \epsilon ^{2}\mu ,\epsilon ^{2}\sigma ^{2}\right) $, and thus it is
necessary to understand our choice of scaling.\footnote{%
Our scaling may bring to mind the small risks modeled by Brownian motion for
which both drift and variance are proportional to the time interval $dt$
(identified here with $\epsilon ^{2}$).} The latter scaling fits and "works
in" our model because the scale-invariant mean-variance ratio matches the
key hypothesis embedded in (\ref{Un}), that as $n$ increases and the payoff
at each stage is effectively a smaller gamble, neither the mean or the
variance dominates.

\subsection{Optimization and the value of a set of arms}

Given a horizon of length $n$, DM solves the following optimization problem: 
\begin{equation}
V_{n}\equiv \sup_{\theta \in \Theta }E_{P}U_{n}\left( \theta \right) \text{.}
\label{Vn}
\end{equation}%
The finite horizon problem is generally not tractable, even when $u$ has the
special form (u.1). For reasons of tractability, Bayesian models in the
literature typically take $\varphi $ in (u.1) to be linear, reducing the
problem to maximization of expected total rewards, but at the cost of
assuming risk neutrality.{\scriptsize \ }Instead, we consider large horizons
and approximate optimality. Then we can accommodate a much more general
class of utility indices.

The first step in developing asymptotics is to define 
\begin{equation}
V\equiv \lim_{n\rightarrow \infty }V_{n}\text{.}  \label{V}
\end{equation}%
Our first theorem proves that $V$ is well-defined, that is, values have a
limit, and more. (Below $||(x,y)||$ denotes the Euclidean norm.)\newline

\begin{theorem}
\label{thm-bandits}{Let }$u\in C(\mathbb{R}^{2})$\ and let payoffs to the $K$
arms conform to (\ref{musigma}), with {$\underline{\sigma }\geq 0$.} Suppose
further that there exists $g\geq 1$ such that $u$ satisfies the growth
condition{\small \ }$|u(x,y)|\leq c(1+||(x,y)||^{g-1})${\small , }and that
payoffs satisfy {$\sup_{1\leq k\leq K}E_{P}[|X_{k}|^{g}]<\infty $}. {Then}:%
%
%
%
%
%
%
%
%
%
%
%
%
%
%
%
\end{theorem}

\begin{description}
\item[(i) Values have a limit:] $\lim_{n\rightarrow \infty }V_{n}$ exists.

\item[(ii) Only means and variances matter:] Consider another set of arms,
described by the random payoffs $X_{k}^{\prime }$, $1\leq k\leq K^{\prime }$%
, and denote the corresponding set of mean-variance pairs by $\mathcal{A}%
^{\prime }$ and the corresponding values by $V_{n}^{\prime }$ and $V^{\prime
}$. Let the mean-variance pairs $\left( \mu _{k}^{\prime },\sigma
_{k}^{\prime\, 2}\right) $ be defined by the obvious counterpart of (\ref%
{musigma}). Then 
\begin{equation*}
\mathcal{A}^{\prime }=\mathcal{A}\text{ }\Longrightarrow ~V^{\prime }=V\text{%
.}
\end{equation*}%
Thus we can write 
\begin{equation*}
V=V\left( \mathcal{A}\text{ }\right) =V\left( \{\left( \mu _{k},\sigma
_{k}^{2}\right) :1\leq k\leq K\}\right) \text{.}
\end{equation*}

\item[(iii) Extreme arms are enough:] 
\begin{equation}
V\left( \mathcal{A}\text{ }\right) =V\left( \mathcal{A}^{ext}\right) \text{.}
\label{Vbound}
\end{equation}%
\bigskip
\end{description}

\noindent \textbf{Remark:} The assumption that $u$ is continuous rules out
example (u.4). However, because these functions can be approximated by
continuous functions, the CLT (Proposition \ref{propn-CLT}) and subsequently
the above theorem, can be extended to cover them as well. (See Chen, Epstein
and Zhang (2023, section A.3), for a similar extension from continuous
functions to indicators.) Similarly for results below. Because the details
are standard, we will ignore the discontinuity of (u.4).

\smallskip

The Appendix contains a proof of (i) and also gives two alternative\textsl{\ 
}expressions for the limit $V$. (ii) describes a simplification for the
decision-maker afforded by adoption of the infinite-horizon heuristic - 
\textit{she need only know and take into account the means and variances for
each arm}. In addition, it permits identifying an arm with its mean-variance
pair; thus we will often refer to a pair $\left( \mu ,\sigma ^{2}\right) $
as an arm. (iii) describes a further possible simplification for DM -- she
need only consider "extreme arms", that is, the extreme points of the convex
polygon generated by $\mathcal{A}$. All other arms are redundant. For
example, \textit{given two arms }$\left( \mu _{1},\sigma _{1}^{2}\right) $%
\textit{\ and }$\left( \mu _{2},\sigma _{2}^{2}\right) $\textit{, then any
arm lying on the straight line between them{\tiny \ }has no value
asymptotically even if it moderates large differences in the mean-variance
characteristics of the two given arms}. For another implication of (iii),
because $\mathcal{A}$ is contained in the rectangle defined by the four
pairs on the right, one obtains that%
\begin{equation*}
V\left( \mathcal{A}\right) \leq V\left( \left\{ (\overline{\mu },\overline{%
\sigma }^{2}),(\overline{\mu },\underline{\sigma }^{2}),(\underline{\mu },%
\overline{\sigma }^{2}),(\underline{\mu },\underline{\sigma }^{2})\right\}
\right) \text{.}
\end{equation*}
Finally, note that both (ii) and (iii) are \textit{true under weak
(nonparametric) assumptions on }$u$\textit{, for example, without any
assumptions about monotonicity or risk attitudes. Therefore,\ they
accommodate situations that feature targets, aspiration levels, loss
aversion, and other deviations from the common assumption of global
monotonicity and risk aversion}.%

The sufficiency of means and variances might be expected from the classic
CLT, and arises here for similar reasons.\footnote{%
This is not to say that the result can be derived from the classical CLT, or
that it is in any way "obvious." Its proof is decidedly nontrivial.} We turn
to intuition for (iii). Consider the evaluation of arm $k$ in the context of
making the contingent decision for stage $i$. If the horizon $n$ is large,
then the payoff to arm $k$ contributes little to the averages determining
overall utility. Accordingly, a second-order Taylor series expansion
provides a good approximation to the incremental benefit from arm $k$, which
expansion, to order $O\left( n^{-1}\right) $, is linear in $\left( \mu
_{k},\sigma _{k}^{2}\right) $. Therefore, the value when maximizing over the 
$K$ arms (asymptotically) equals that when maximizing over the convex hull
of $\mathcal{A}$, or over its set of extreme points $\mathcal{A}^{ext}$, as
asserted in (\ref{Vbound}). \textit{In more economic terms, extreme arms are
sufficient because switching suitably between them across stages can, in the
infinite-horizon limit, replicate or improve upon the payoff distribution
achievable when choosing from the entire set of }$K$\textit{\ arms.} \ 

\subsection{Strategies and the risk/reward tradeoff}

Turn to strategies. Given the $K$ arms corresponding to $\mathcal{A}$, the
strategy $\theta ^{\ast }$ is \emph{asymptotically optimal} if 
\begin{equation*}
\lim_{n\rightarrow \infty }E_{P}U_{n}\left( \theta ^{\ast }\right) =V\left( 
\mathcal{A}\right) \text{.}
\end{equation*}%
It follows that $\theta ^{\ast }$ is \emph{approximately optimal} for large
horizons in that: for every $\epsilon >0$, there exists $n^{\ast }$ such
that 
\begin{equation*}
\mid U_{n}\left( \theta ^{\ast }\right) -V_{n}\mid <\epsilon \text{ \ if }%
n>n^{\ast }\text{.}
\end{equation*}

Say that $\left( \mu ,\sigma ^{2}\right) $ is \emph{feasible} if it lies in $%
\mathcal{A}$. Theorem \ref{thm-bandits}(iii) states that DM can limit
herself to strategies that choose between extreme arms. More can be said
under added assumptions on the utility index and what is feasible, as
illustrated by the next result.

\begin{theorem}
\label{thm-strategies}Adopt the assumptions in Theorem \ref{thm-bandits}. If 
$u(x,y)$ is increasing in $x$ and concave in $y$, and if $(\overline{\mu },%
\underline{\sigma }^{2})$ is feasible, then: the strategy of always choosing
an arm exhibiting $(\overline{\mu },\underline{\sigma }^{2})$ is
asymptotically optimal, and the corresponding limiting value, defined in (%
\ref{V}), is given by 
\begin{equation*}
V=\int u\left( \overline{\mu },\cdot \right) d\mathbb{N}\left( 0,\underline{%
\sigma }^{2}\right) \text{.}
\end{equation*}
\end{theorem}

Intuition argues for the choice of $\left( \overline{\mu },\underline{\sigma 
}^{2}\right) $ at stage $n$ if there are no later trials remaining, but may
seem myopic more generally. Notably, the strategy of always choosing the
high-mean/low-variance pair is not in general optimal given a finite horizon
(even apart from the fact that arms may not be adequately characterized by
mean and variance alone). That it is asymptotically optimal demonstrates a
simplifying feature of the long-horizon heuristic. An additional comment is
that one can similarly consider three other possible combinations of
monotonicity and curvature assumptions for $u$, where each property is
assumed to hold globally. For example, if $u(x,y)$ is decreasing in $x$ and
concave (convex) in $y$, then it is asymptotically optimal to always choose
an arm exhibiting $(\underline{\mu },\underline{\sigma }^{2})$ ($(\underline{%
\mu },\overline{\sigma }^{2})$) if it is feasible.

However, the theorem does not provide any insight into the risk/reward
tradeoff that is at the core of decision-making under uncertainty. Under
common assumptions about monotonicity and risk aversion, the tradeoff
concerns the increase in mean reward needed to compensate the individual for
facing an increase in risk (for example, a larger variance). But Theorem \ref%
{thm-strategies} assumes that there exists an arm having \textsl{both} the
largest mean \textsl{and} the smallest variance, thus ruling out the need
for DM to make such a tradeoff.

Next we investigate asymptotic optimality when the risk/reward tradeoff is
integral. For greater clarity, we do so in a canonical setting where there
are 2 arms ($K=2$), described by $\left( \mu _{1},\sigma _{1}^{2}\right) $\
and $\left( \mu _{2},\sigma _{2}^{2}\right) $, and where 
\begin{equation}
\mu _{1}>\mu _{2}\text{, }\ \sigma _{1}>\sigma _{2}\geq 0\text{.}
\label{mv2}
\end{equation}%
Parts (i) and (ii) of the next theorem describe conditions under which it is
asymptotically optimal to \emph{specialize }in one arm, that is, to choose
that arm always (at every stage and history). The remaining parts give
conditions under which specializing in one arm is not asymptotically optimal
(that is, not even approximately optimal for large horizons). Some results
are limited to utility specifications in the Example.

\begin{theorem}
\label{thm-tradeoff} Adopt the assumptions in Theorem \ref{thm-bandits} and
consider the 2-arm case above. Then, for each of the following
specifications of $u$, the indicated strategy is asymptotically optimal and $%
V$ denotes the corresponding limiting value defined in (\ref{V}).
\end{theorem}

\begin{description}
\item[(i)] Let $u:\mathbb{R}^{2}\longrightarrow \mathbb{R}$ be twice
continuously differentiable. Suppose that 
\begin{equation}
\partial _{x}u\left( x,y\right) \left( \mu _{1}-\mu _{2}\right) +\tfrac{1}{2}%
\partial _{yy}^{2}u\left( x,y\right) \left( \sigma _{1}^{2}-\sigma
_{2}^{2}\right) \geq 0\text{ \ for all }(x,y)\in \mathbb{R}^{2}.\newline
\label{AAi}
\end{equation}%
Then specializing in arm 1 always is asymptotically optimal and, (by (\ref%
{Uiid})), $V=\int u\left( \mu _{1},\cdot \right) d\mathbb{N}\left( 0,\sigma
_{1}^{2}\right) $. If $\partial _{x}u$ is everywhere positive, then (\ref%
{AAi}) is equivalent to 
\begin{equation}
\frac{-\frac{1}{2}\partial _{yy}^{2}u\left( x,y\right) }{\partial
_{x}u\left( x,y\right) }\leq \frac{\mu _{1}-\mu _{2}}{\sigma _{1}^{2}-\sigma
_{2}^{2}}\text{ for all }(x,y)\in \mathbb{R}^{2}\text{.}  \label{AAi'}
\end{equation}%
When the inequality in (\ref{AAi}) is reversed, then it is asymptotically
optimal to specialize in arm 2.

\item[(ii)] Adopt the conditions on $u$ in (i), and assume that $\partial
_{x}u\left( x,y\right) >0$ for all $\left( x,y\right) \in \mathbb{R}^{2}$.
Suppose further that 
\begin{equation}
\frac{-\frac{1}{2}\partial _{yy}^{2}u}{\partial _{x}u}=\alpha >0\text{ ~for
all }\left( x,y\right) \in \mathbb{R}^{2}\text{.}  \label{constAA}
\end{equation}%
Then specializing in arm $1$ (arm $2$) is asymptotically optimal if 
\begin{equation}
\alpha \leq \text{(}\geq \text{) }\frac{\mu _{1}-\mu _{2}}{\sigma
_{1}^{2}-\sigma _{2}^{2}}\text{.}  \label{AAii}
\end{equation}%
Both strategies are asymptotically optimal when there is equality in (\ref%
{AAii}).

\item[(iii)] Let $u\left( x,y\right) =x-\alpha y^{2}I_{(-\infty ,0)}\left(
y\right) ,\alpha >0$. Observe that 
\begin{equation*}
\frac{\mu _{1}-\mu _{2}}{\sigma _{1}^{2}-\sigma _{2}^{2}}<\underline{\alpha }%
<\overline{\alpha }\text{,}
\end{equation*}%
where the critical values $\underline{\alpha }$ and $\overline{\alpha }$ are
given by\footnote{$\overline{\alpha }=\infty $ if $\sigma _{2}=0$.}%
\begin{equation*}
\underline{\alpha }\equiv \frac{4(\mu _{1}-\mu _{2})}{3(\sigma
_{1}^{2}-\sigma _{2}^{2})}\text{, }\overline{\alpha }\equiv \frac{2(\mu
_{1}-\mu _{2})}{\sigma _{2}(\sigma _{1}-\sigma _{2})}\text{.}
\end{equation*}%
If $\alpha \leq \frac{\mu _{1}-\mu _{2}}{\sigma _{1}^{2}-\sigma _{2}^{2}}$,
then specializing in arm $1$ is asymptotically optimal. If $\underline{%
\alpha }<\alpha $ (respectively $\alpha <\overline{\alpha }$), then
specializing in arm 1 (arm 2) is \emph{not} asymptotically optimal, from
which it follows that specialization in \emph{either} arm is \emph{not}
asymptotically optimal if $\underline{\alpha }<\alpha <\overline{\alpha }$.

\item[(iv)] Let $u\left( x,y\right) =\noindent \ x-\alpha I_{(-\infty
,0)}\left( y\right) $, $\alpha >0$. Let $\sigma _{2}>0$. Then specializing
in arm 2 is \emph{not} asymptotically optimal for any $\alpha $, and, if 
\begin{equation*}
\underline{\alpha }^{\prime }\equiv \frac{2(\mu _{1}-\mu _{2})\sigma _{1}}{%
\left( \sigma _{1}-\sigma _{2}\right) }<\alpha \text{,}
\end{equation*}%
then neither is specializing in arm 1.

\item[(v)] Let $u\left( x,y\right) =\varphi \left( x\right) +\alpha y$, $%
\varphi \in C\left( \mathbb{R}\right) $ and $\alpha \in \mathbb{R}$. {Fix $%
x^{\ast }\in \arg \max\limits_{\mu _{1}\leq x\leq \mu _{2}}\varphi (x)$, and
let $\lambda \in \lbrack 0,1]$ be such that $x^{\ast }=\lambda \mu
_{1}+(1-\lambda )\mu _{2}$. Denote by }$\psi _{i}${\ the number times that
arm $1\ $is chosen in first }$i${\ stages. Let the strategy $\theta ^{\ast }$
choose arm }$1${\ at stage $1$, and also at stage $i+1$,$\ $($i\geq 1$), if
and only if $\tfrac{\psi _{i}}{i}\leq \lambda $. Then }$\theta ^{\ast }$ is
asymptotically optimal and{%
\begin{equation*}
V={\max\limits_{\mu _{2}\leq x\leq \mu _{1}}\varphi (x).}
\end{equation*}%
} 
Further, specializing in one arm is asymptotically optimal if and only if 
\newline
$\max \{\varphi \left( \mu _{1}\right) ,\varphi \left( \mu _{2}\right) \}={%
\max\limits_{\mu _{2}\leq x\leq \mu _{1}}\varphi (x)}$.

\item \bigskip
\end{description}

\noindent \textbf{Remark}: It is straightforward to extend the theorem to an
arbitrary set of $K$ arms. For example, in (i), with $\partial _{x}u$
everywhere positive, specializing in arm $j$\ is asymptotically optimal if%
\textit{\ }%
\begin{equation*}
j\in \arg \max_{k=1,...,K}\{\mu _{k}-(\tfrac{-\frac{1}{2}\partial
_{yy}^{2}u\left( x,y\right) }{\partial _{x}u})\sigma _{k}^{2}\}\text{ ~for
all }\left( x,y\right) \text{,}
\end{equation*}%
which simplifies in the obvious way under the constancy condition (\ref%
{constAA}).

\bigskip

We discuss each part of the theorem in turn.

\medskip

(i) Focus on (\ref{AAi'}). Intuition derives from interpretation given above
of $-\partial _{yy}^{2}u/\partial _{x}u$ as a (local) measure of risk
aversion.%
%
%
The relatively small degree of risk aversion indicated in (\ref{AAi'})
implies that the larger mean for arm 1 more than compensates for its larger
variance. Moreover, this is true at each stage, regardless of history,
because the inequality in (\ref{AAi'}) is satisfied globally.

\medskip

(ii) This is an immediate consequence of (i) that we include in the
statement because the consequence of the indicated constancy warrants
emphasis. Two examples covered by this constancy are mean-variance and the
special case of (u.2) where $\varphi $ is an exponential.%
%
%
At first glance, the implication regarding the \textit{unimportance of
diversification} might seem surprising, especially given its central role in
portfolio theory. Of course, diversification in portfolio theory refers to
the simultaneous holding of several assets, which, interpreting each arm as
an asset, is excluded here. But diversification over time is permitted and
that is its meaning here. The result that specialization in one arm over
time is always asymptotically optimal given (\ref{constAA}) can be
understood as follows. Considering the factors that might lead to different
arms being chosen at two different stages, note first that the payoff
distribution for each arm is unchanged by assumption. Second, though a
finite-horizon induces a nonstationarity that can affect choices, our
decision-maker is, roughly speaking, acting as if solving an
infinite-horizon problem. That leaves only the variation of risk attitude
with past outcomes, which is excluded if $-\partial _{yy}^{2}u/\partial
_{x}u $ is constant.

\medskip

(iii) Note first that it has often been argued, including by Markowitz
(1959), that investors are more concerned with downside risk than with
variance, and hence that \textit{semivariance is a better measure of the
relevant risk}. In our sequential choice context, the mean-semivariance
model agrees partially with the mean-variance model in that for both (the
inequality $\leq $ in) (\ref{AAii}) implies the asymptotic optimality of
choosing (the high mean, high variance) arm 1 throughout. However, their
agreement ends there. In particular, there is \textit{a role for
time-diversification} for the semivariance model, in that, for $\underline{%
\alpha }<\alpha <\overline{\alpha }$, asymptotic optimality can be achieved
only by a strategy that employs both arms. (In particular, if arm 2 is
risk-free ($\sigma _{2}=0$), then time-diversification is necessary for
asymptotic optimality if $\frac{3}{4}\alpha $ exceeds the risk-adjusted
excess mean $(\mu _{1}-\mu _{2})/\sigma _{1}^{2}$.) Here is some intuition
for the existence of a region with nonspecialization. Since only negative
deviations are penalized, it is as though DM faces, or perceives, less risk
than what is measured by $\sigma ^{2}$. Alternatively, in our preferred
interpretation, for any given risk measured by variance, DM is less averse
to that risk in the present model as if her effective $\alpha $ is smaller
than its nominal magnitude. Moreover, risk aversion varies across stages.
For example, contingent on cumulative past deviations being positive
(negative) at stage $m$,$\ $it is relatively unlikely (likely) that future
choices will lead later to negative cumulative deviations, and thus variance
is less (more) of a concern. Such \textit{endogenous changes in risk aversion%
} can lead to specialization in either single arm being dominated in large
horizons.

In finance, it has been argued (Nantell and Price 1979; Klebaner et al 2017)
that the change from variance to semivariance has limited consequences for
received asset market theory. In contrast, a similar change in the bandit
problem context leads to qualitative differences regarding the importance of
time-diversification.{\scriptsize \ }

\medskip

(iv) This utility specification, for which only the existence of a shortfall
and not its size matters, implies that \textit{it is never asymptotically
optimal to specialize in the low mean, low variance arm}.\footnote{%
Intuitively, relying exclusively on the more conservative arm increases the
asymptotic likelihood of cumulative shortfalls. The problem is reminiscent
of the classic introductory story in Dubins and Savage (1976, Ch. 1) of a
gambler who must decide how to gamble in order to minimize the probability
that cumulative winnings fall short of a fixed target. Their solution is
that he should not gamble cautiously.} Indeed, by (\ref{u5limit}),
specializing in the high mean, high variance arm is superior for large
horizons without any regard to the numerical magnitudes of $\mu _{1}-\mu
_{2} $ and $\sigma _{1}^{2}-\sigma _{2}^{2}$. However, specializing in the
high mean, high variance arm is also ruled out for large enough $\alpha $ -
those lying in $\left( \underline{\alpha }^{\prime },\infty \right) $. Note
that this set grows larger as $\sigma _{1}$ increases (keeping $\mu _{1}$, $%
\mu _{2}$ and $\sigma _{2}$ fixed) - a larger variance makes it more likely
that repeated choice of arm 1 will produce a cumulative shortfall, which is
tolerable only if the associated penalty parameter $\alpha $ is smaller.
Therefore, as in the semivariance model (iii), for a range of parameter
values \textit{asymptotic optimality can be achieved only through
diversification across time}.

\medskip

(v) The utility specification $u\left( x,y\right) =\varphi \left( x\right)
+\alpha y$ leads to an asymptotically optimal strategy that is diversified
and that can be described explicitly. Condition (\ref{AAi}) suggests that
either nonmonotonicity (e.g. a change in the sign of $\partial _{x}u$), or
variable risk aversion (e.g. a change in the sign or magnitude of $\partial
_{yy}^{2}u$) might lead to the asymptotic optimality of switching between
arms. This utility specification, with $\varphi $ not necessarily monotonic,
illustrates the former case. The interpretation of the strategy $\theta
^{\ast }$ defined in the theorem is that \textit{DM targets }$x^{\ast }$%
\textit{, a maximizer of }$\varphi $ on $\left[ \mu _{2},\mu _{1}\right] $.
(When $(\mu _{2}+\mu _{1})/2$ is a maximizer, then $\theta ^{\ast }$ chooses
arms according to the sequence $121212...$. When $\varphi $ is monotonic, $%
\theta ^{\ast }$ specializes in arm 1 or in arm 2 according as $\varphi $ is
increasing or decreasing on $\left[ \mu _{2},\mu _{1}\right] $,
respectively.) Irrespective of any nonlinearity of $\varphi $, and the
implied non-neutrality to risk, \textit{variances do not matter
asymptotically as in the classic LLN}.

\section{Concluding Comments}

Our model has produced new results regarding sequential choice between
repeated gambles, most notably in describing connections, expressed in
simple formal terms, between the endogeneity of risk aversion and the value
of time-diversification. Three features of the model that facilitate
tractability are (i) the heuristic of approximate optimality for large
horizons, which is the decision-maker's assumed response to a complex
problem; (ii) the existence of a suitable measure of risk attitude (similar
to, but distinct from, the Arrow-Pratt measure) that describes her
risk/reward tradeoff; and (iii) the fact that without loss of generality
gambles can be represented by their mean and variance alone, thus providing
a new rationale for mean-variance analysis.

The results are general in the sense that payoff distributions are
unrestricted except for the requirement that means and variances exist.
However, results depend on our nonstandard specification of payoffs (via
averages) and utility function. Specific assumptions are needed in order to
derive analytical results, and our assumption compares favourably, in our
view, with the assumption of risk neutrality adopted in much of the bandit
literature. Given the complexity of dynamic decision problems under
uncertainty, it is natural to wonder if behavior might be better described
by "approximate optimality," and we view the paper as a modest first step in
this modeling direction. Undoubtedly more needs to be done. Axiomatic
analysis for such behavior poses an interesting challenge for decision
theorists.

\appendix

\section{Appendix: Proofs\label{section-proofs}}

We remind the reader of the following notation used in this section: $%
\overline{\mu },\underline{\mu }$ and $\overline{\sigma }^{2},\underline{%
\sigma }^{2}$ are the bounds of means and variances given in (\ref{bars}), $%
\mathcal{A}$ denotes the set of mean-variance pairs of all $K$ arms, and $%
\mathcal{A}^{ext}\subset \mathcal{A}$ denotes the set of extreme points of $%
co\left( \mathcal{A}\right) $. Pairs consisting of mean and standard
deviation (rather than variance) will also be important, and thus it is
convenient to define 
\begin{eqnarray*}
\lbrack \mathcal{A}] &=&\{\left( \mu ,\sigma \right) :\left( \mu ,\sigma
^{2}\right) \in \mathcal{A}\}\text{, and} \\
\lbrack \mathcal{A}]^{ext} &=&\{\left( \mu ,\sigma \right) :\left( \mu
,\sigma ^{2}\right) \in \mathcal{A}^{ext}\}
\end{eqnarray*}

{Let $B=\{B_{t}=(B_{t}^{(1)},B_{t}^{(2)})\}$ be a two-dimensional standard
Brownian motion defined on $(\Omega ,\mathcal{F},P)$, and let $\{\mathcal{F}%
_{t}\}$ be the natural filtration generated by $(B_{t})$. For a fixed $T>0$,
and any $0\leq t\leq s\leq T$, 
let $[\mathcal{A}](t,T)$ denote the set of all }$\{\mathcal{F}_{s}\}$%
-progressively measurable processes, $a=\{a_{s}=(a_{s}^{(1)},a_{s}^{(2)})%
\}:[t,T]\times \Omega \rightarrow \lbrack \mathcal{A}]\subset \mathbb{R}^{2}$%
. Finally, $[${$\mathcal{A}]^{ext}(t,T)$ is }defined similarly by
restricting the images of each process $a$ to lie in $[${$\mathcal{A}]^{ext}$%
. }


The following lemma gives properties of $\{Z_{n}^{\theta }\}$ that will be
used repeatedly.

\begin{lemma}
\label{tab-proper} The rewards $\{Z_{n}^{\theta }:n\geq 1\}$ defined in (\ref%
{zntheta}) satisfy the following:

\begin{description}
\item[(1)] For any $n\geq 1$, 
\begin{align*}
\overline{\mu }& =ess\sup\limits_{\theta \in \Theta }E_{P}[Z_{n}^{\theta }|%
\mathcal{H}_{n-1}^{\theta }],\ \underline{\mu }=ess\inf\limits_{\theta \in
\Theta }E_{P}[Z_{n}^{\theta }|\mathcal{H}_{n-1}^{\theta }] \\
\overline{\sigma }^{2}& =ess\sup\limits_{\theta \in \Theta }E_{P}\left[
\left( Z_{n}^{\theta }-E_{P}[Z_{n}^{\theta }|\mathcal{H}_{n-1}^{\theta
}]\right) ^{2}|\mathcal{H}_{n-1}^{\theta }\right] \\
\underline{\sigma }^{2}& =ess\inf\limits_{\theta \in \Theta }E_{P}\left[
\left( Z_{n}^{\theta }-E_{P}[Z_{n}^{\theta }|\mathcal{H}_{n-1}^{\theta
}]\right) ^{2}|\mathcal{H}_{n-1}^{\theta }\right] .
\end{align*}

\item[(2)] For any $\theta \in \Theta $ and $n\geq 1$, let $U_{n-1}^{\theta
} $ be any $\mathcal{H}_{n-1}^{\theta }$-measurable random variable. For any
bounded measurable functions $f_{0},f_{1}$ and $f_{2}$ on $\mathbb{R}$, let $%
\psi (x,y)=f_{0}(x)+f_{1}(x)y+f_{2}(x)y^{2},(x,y)\in \mathbb{R}^{2}$. Then%
{\small 
\begin{equation*}
\sup_{\theta \in \Theta }E_{P}\left[ \psi \left( U_{n-1}^{\theta
},Z_{n}^{\theta }\right) \right] =\sup\limits_{\theta \in \Theta }E_{P}\left[
\max_{1\leq k\leq K}\left\{ \psi _{k}\big(U_{n-1}^{\theta }\big)\right\} %
\right]
\end{equation*}%
}where, for all $x\in ${\normalsize $\mathbb{R}$ }and $1\leq k\leq K$, 
\begin{equation}
\psi _{k}(x)=E_{P}[\psi (x,X_{k,n})]=f_{0}(x)+\mu _{k}\,f_{1}(x)+(\mu
_{k}^{2}+\sigma _{k}^{2})\,f_{2}(x).  \label{psi-nk}
\end{equation}
\end{description}
\end{lemma}

\textbf{Proof:} (1) $\{Z_{n}^{\theta }\}$ satisfy, for any $\theta \in
\Theta $ and $n\geq 1$, 
\begin{align*}
E_{P}[Z_{n}^{\theta }|\mathcal{H}_{n-1}^{\theta }]=&
\sum_{k=1}^{K}I_{\{\theta _{n}=k\}}E_{P}[X_{k,n}|\mathcal{H}_{n-1}^{\theta }]
\\
=& \sum_{k=1}^{K}I_{\{\theta
_{n}=k\}}E_{P}[X_{k,n}]=\sum_{k=1}^{K}I_{\{\theta _{n}=k\}}\mu _{k}.
\end{align*}%
Combine with the definitions of $\overline{\mu }$ and $\underline{\mu }$ in (%
\ref{bars}) to derive 
\begin{equation*}
ess\sup_{\theta \in \Theta }E_{P}[Z_{n}^{\theta }|\mathcal{H}_{n-1}^{\theta
}]=\overline{\mu },\ ess\inf_{\theta \in \Theta }E_{P}[Z_{n}^{\theta }|%
\mathcal{H}_{n-1}^{\theta }]=\underline{\mu }.
\end{equation*}%
The other two equalities can be proven similarly.\vspace{0.2cm}

(2) For any $\theta \in \Theta $ and $n\geq 1$, let $U_{n-1}^{\theta }$ be a 
$\mathcal{H}_{n-1}^{\theta }$-measurable random variable. By direct
calculation we obtain that%
\begin{align*}
& \sup\limits_{\theta \in \Theta }E_{P}\left[ \psi \left( U_{n-1}^{\theta
},Z_{n}^{\theta }\right) \right] \\
=& \sup\limits_{\theta \in \Theta }E_{P}\left[ \sum_{k=1}^{K}I_{\{\theta
_{n}=k\}}E_{P}[\psi \left( U_{n-1}^{\theta },X_{k,n}\right) |\mathcal{H}%
_{n-1}^{\theta }]\right] \\
=& \sup\limits_{\theta \in \Theta }E_{P}\left[ \max_{1\leq k\leq K}\psi _{k}%
\big(U_{n-1}^{\theta }\big)\right] ,
\end{align*}%
where $\psi _{k}$ is given in (\ref{psi-nk}). {\normalsize \hfill $%
\blacksquare $}

\medskip

Following Peng (2019), our arguments make use of nonlinear partial
differential equations (PDEs) and viscosity solutions. The following is
taken from {Theorems 2.1.2, C.3.4 and C.4.5} in Peng's book.


\begin{lemma}
\label{lemma-pde} For given $T>0$, consider the following PDE: 
\begin{equation}
\left\{ 
\begin{array}{l}
\partial _{t}v(t,x,y)+G\left( \partial _{x}v(t,x,y),\partial
_{yy}^{2}v(t,x,y)\right) =0,\quad (t,x,y)\in \lbrack 0,T)\times \mathbb{R}%
^{2} \\ 
v(T,x,y)=u(x,y),%
\end{array}%
\right.  \label{pde-1}
\end{equation}%
where $u\in C${\normalsize $(\mathbb{R}^{2})$. }Suppose that $G$ is
continuous on {\normalsize $\mathbb{R}^{2}$ }and satisfies the following
conditions, for all {\normalsize $(p,q),(p^{\prime },q^{\prime })\in \mathbb{%
R}^{2}$}: 
\begin{align}
G(p,q)& \leq G(p,q^{\prime }),\ \text{ whenever }q\leq q^{\prime },
\label{G-condition-1} \\
G(p,q)-G(p^{\prime },q^{\prime })& \leq G(p-p^{\prime },q-q^{\prime }),
\label{G-condition-2} \\
G(\lambda p,\lambda q)& =\lambda G(p,q),\ \text{ for }\lambda \geq 0.
\label{G-condition-3}
\end{align}%
Then, for any $u\in C(\mathbb{R}^{2})$ satisfying a polynomial growth
condition, there exists a unique $v\in C([0,T]\times \mathbb{R}^{2})$ such
that $v$ is a viscosity solution of the PDE (\ref{pde-1}). Moreover, if $%
\exists \lambda >0$ such that, for all $p\in \mathbb{R}$ \ and $q\geq
q^{\prime }\in \mathbb{R}$, 
\begin{equation*}
G(p,q)-G(p,q^{\prime })\geq \lambda (q-q^{\prime }),
\end{equation*}%
and if the initial condition $u$ is uniformly bounded, then for each $%
0<\epsilon <T$, $\exists \beta \in (0,1)$ such that 
\begin{equation}
\Vert v\Vert _{C^{1+\beta /2,2+\beta }([0,T-\epsilon ]\times \mathbb{R}%
^{2})}<\infty .  \label{smooth}
\end{equation}%
Here $\Vert \cdot \Vert _{C^{1+\beta /2,2+\beta }([0,T-\epsilon ]\times 
\mathbb{R}^{2})}$ is the Krylov (1987) norm on \newline
$C^{1+\beta /2,2+\beta }([0,T-\epsilon ]\times \mathbb{R}^{2})$, the set of
(continuous and) suitably differentiable functions on $[0,T-\epsilon ]\times 
\mathbb{R}^{2}$.\footnote{%
Some detail is provided in the Online Appendix. See also Peng (2019, Ch.
2.1).}
\end{lemma}

\subsection{Proof of Theorem \protect\ref{thm-bandits}}


We first prove a nonlinear central limit theorem for the bandit problem. The
values $V_{n}$ and $V$ are defined in (\ref{Vn}) and (\ref{V}) respectively.

\begin{proposition}[CLT]
\label{propn-CLT}Let $u\in C_{b,Lip}(\mathbb{R}^{2})$, {the class of all
bounded and Lipschitz continuous functions on $\mathbb{R}^{2}$, and }adopt
all other assumptions and the notation in Theorem \ref{thm-bandits}. Then 
\begin{eqnarray}
\lim_{n\rightarrow \infty }V_{n} &=&V=\sup_{a\in \lbrack \mathcal{A}%
](0,1)}E_{P}\left[ u\left(
\int_{0}^{1}a_{s}^{(1)}ds,\int_{0}^{1}a_{s}^{(2)}dB_{s}^{(2)}\right) \right]
\label{lemma-CLT-eq} \\
&=&\sup_{a\in \lbrack \mathcal{A}]^{ext}(0,1)}E_{P}\left[ u\left(
\int_{0}^{1}a_{s}^{(1)}ds,\int_{0}^{1}a_{s}^{(2)}dB_{s}^{(2)}\right) \right]
.  \label{lemma-CLT-eq-extreme}
\end{eqnarray}
\end{proposition}

The proof in this appendix assumes \underline{$\sigma $}$>0$. The extension
to \underline{$\sigma $}$=0$ is proven in the Online Appendix. The
boundedness assumption on utility indices excludes many interesting
specifications. However, the Online Appendix shows that the Proposition is
valid for all $u\in C\left( \mathbb{R}^{2}\right) $ satisfying a growth
condition.

The following immediate corollary is used frequently in later proofs of
Theorems \ref{thm-strategies} and \ref{thm-tradeoff} (the Online Appendix
contains a proof).

\begin{corollary}
\label{cor-CLTlimit}For all $u\in C\left( \mathbb{R}^{2}\right) $ satisfying
a polynomial growth condition, the limit in (\ref{lemma-CLT-eq}) can be
described also by the solution of a PDE. Specifically, 
\begin{equation}
V=v(0,0,0),  \label{Vv}
\end{equation}%
where $v$ is the solution of PDE (\ref{pde-1}), 
with function $G$ given by%
\begin{equation}
G(p,q)=\sup_{(\mu ,\sigma ^{2})\in \mathcal{A}}\left[ \mu p+\tfrac{1}{2}%
\sigma ^{2}q\right] ,\quad (p,q)\in \mathbb{R}^{2}.  \label{G-function}
\end{equation}
\end{corollary}

Some related literature on CLTs was cited in the introduction. In addition,
Chen and Epstein (2022) and Chen, Epstein and Zhang (2022) have nonlinear
CLTs, which, when translated into the bandits context, restrict differences
between arms either by assuming that they all have the identical variance
(in the former paper), or the identical mean (in the latter paper). These
restrictions preclude study of the risk/reward tradeoff. In addition, their
objective is to obtain simple closed-form expressions for the limit (what we
denote by $V$), and for that purpose they adopt very special functional
forms for $u$.\footnote{%
In particular, the second paper cited assumes $u\left( x,y\right) =\varphi
\left( y\right) $, where $\varphi \left( y\right) =\varphi _{1}\left(
y-c\right) $ if $y\geq c$, and $=-\lambda ^{-1}\varphi _{1}\left( -\lambda
(y-c)\right) $ if $y<c$, for some function $\varphi _{1}$ and $c\in \mathbb{R%
}$. This functional form is motivated by loss aversion, but from the
perspective of this paper is very special.} In contrast, Proposition \ref%
{propn-CLT} and its corollary apply to a much more general class of utility
indices. Moreover, as this paper shows, in spite of the complexity of the
expression for $V$ it is the basis for a range of results about the bandit
problem even allowing unrestricted heterogeneity across arms.

Next we proceed with lemmas that will lead to a proof of the CLT.\textsl{\ }%
They assume $u\in C_{b}^{3}(\mathbb{R}^{2})$ and relate to the functions $%
\{H_{t}\}_{t\in \lbrack 0,1]}$ defined by, for all $\ (x,y)\in \mathbb{R}%
^{2} $, 
\begin{equation}
H_{t}(x,y)=\sup_{a\in \lbrack \mathcal{A}](t,1+h)}E_{P}\left[ u\left(
x+\int_{t}^{1+h}a_{s}^{(1)}ds,y+\int_{t}^{1+h}a_{s}^{(2)}dB_{s}^{(2)}\right) %
\right] ,  \label{Ht}
\end{equation}%
where $h>0$ is fixed and dependence on $h$ is suppressed notationally. In
addition, we often write $z=\left( z_{1},z_{2}\right) =\left( x,y\right) $
and define $|z-z^{\prime }|^{\beta }=$ $|z_{1}-z_{1}^{\prime }|^{\beta
}+|z_{2}-z_{2}^{\prime }|^{\beta }$.\bigskip

\begin{lemma}
\label{lemma-ddp-G}The functions $\{H_{t}\}_{t\in \lbrack 0,1]}$ satisfy the
following properties:

\begin{description}
\item[(1)] $H_{t}\in C_{b}^{2}(\mathbb{R}^{2})$ and the first and second
derivatives of $H_{t}$ are uniformly bounded for all $t\in \lbrack 0,1]$.

\item[(2)] There exist constants $L>0$ and $\beta \in (0,1)$, independent of 
$t$, such that for any $(z_{1},z_{2}),(z_{1}^{\prime },z_{2}^{\prime })\in 
\mathbb{R}^{2}$, 
\begin{equation*}
|\partial _{z_{i}z_{j}}^{2}H_{t}(z_{1},z_{2})-\partial
_{z_{i}z_{j}}^{2}H_{t}(z_{1}^{\prime },z_{2}^{\prime })|\leq
L(|z_{1}-z_{1}^{\prime }|^{\beta }+|z_{2}-z_{2}^{\prime }|^{\beta }),\quad
i,j=1,2.
\end{equation*}

\item[(3)] Dynamic programming principle: For any $\delta \in \lbrack
0,1+h-t]$, 
\begin{equation*}
H_{t}\left( x,y\right) =\sup_{a\in \lbrack \mathcal{A}](t,t+\delta )}E_{P}%
\left[ H_{t+\delta }\left( x+\int_{t}^{t+\delta
}a_{s}^{(1)}ds,y+\int_{t}^{t+\delta }a_{s}^{(2)}dB_{s}^{(2)}\right) \right]
,\;(x,y)\in \mathbb{R}^{2}.
\end{equation*}

\item[(4)] For the function $G$ given in (\ref{G-function}), \ we have%
\begin{equation*}
\lim\limits_{n\rightarrow \infty }\sum\limits_{m=1}^{n}\sup\limits_{(x,y)\in 
\mathbb{R}^{2}}\left\vert H_{\frac{m-1}{n}}(x,y)-H_{\frac{m}{n}}\left(
x,y\right) -\frac{1}{n}G\left( {\partial _{x}H_{\frac{m}{n}}\left(
x,y\right) },{\partial _{yy}^{2}H_{\frac{m}{n}}\left( x,y\right) }\right)
\right\vert =0.
\end{equation*}

\item[(5)] There exists a constant $C_{0}>0$ such that%
\begin{equation*}
\sup_{(x,y)\in \mathbb{R}^{2}}|H_{1}(x,y)-u(x,y)|\leq C_{0}h
\end{equation*}%
\begin{equation*}
\sup_{(x,y)\in \mathbb{R}^{2}}|H_{0}(x,y)-\psi (x,y)|\leq C_{0}h,
\end{equation*}%
where {\normalsize $\psi (x,y)=\sup_{a\in \lbrack \mathcal{A}](0,1)}E_{P}%
\left[ u\left(
x+\int_{0}^{1}a_{s}^{(1)}ds,y+\int_{0}^{1}a_{s}^{(2)}dB_{s}^{(2)}\right) %
\right] $. }
\end{description}
\end{lemma}

\noindent {\normalsize \noindent }\textbf{Proof:} For any $t\in \lbrack
0,1+h]$ and {\normalsize $(x,y)\in \mathbb{R}^{2}$, }we define the function $%
v(t,x,y)=H_{t}(x,y)$. 
Then $v$ is the solution of the HJB-equation (\ref{pde-1}) with function $G$
given in (\ref{G-function}) (Yong and Zhou (1999, Theorem 5.2, Ch. 4)).%
\textbf{\ }By Lemma~\ref{lemma-pde}, $\exists \beta \in (0,1)$ such that 
\begin{equation*}
\Vert v\Vert _{C^{1+\beta /2,2+\beta }([0,1]\times \mathbb{R}^{2})}<\infty .
\end{equation*}%
This proves both (1) and (2).

\medskip

\noindent (3) follows directly from the classical dynamic programming
principle ({Yong and Zhou (1999, Theorem 3.3, Ch. 4}))\textbf{.}

\medskip

\noindent Prove (4): 
By Ito's formula, 
\begin{align*}
& \sum_{m=1}^{n}\sup_{(x,y)\in \mathbb{R}^{2}}\left\vert H_{\frac{m-1}{n}%
}(x,y)-H_{\frac{m}{n}}\left( x,y\right) -\frac{1}{n}G\left( {\partial _{x}H_{%
\frac{m}{n}}\left( x,y\right) },{\partial _{yy}^{2}H_{\frac{m}{n}}\left(
x,y\right) }\right) \right\vert \\
=& \sum_{m=1}^{n}\sup_{(x,y)\in \mathbb{R}^{2}}\left\vert \sup_{\alpha \in
\lbrack \mathcal{A}](\frac{m-1}{n},\frac{m}{n})}\!\!E_{P}\left[ H_{\frac{m}{n%
}}\left( x+\int_{\frac{m-1}{n}}^{\frac{m}{n}}a_{s}^{(1)}ds,y+\int_{\frac{m-1%
}{n}}^{\frac{m}{n}}a_{s}^{(2)}dB_{s}^{(2)}\right) \right] \right. \\
& \hspace{3.8cm}-H_{\frac{m}{n}}\left( x,y\right) -\frac{1}{n}G\left( {%
\partial _{x}H_{\frac{m}{n}}\left( x,y\right) },{\partial _{yy}^{2}H_{\frac{m%
}{n}}\left( x,y\right) }\right) \Bigg\vert
\end{align*}%
\begin{align*}
=& \sum_{m=1}^{n}\sup_{(x,y)\in \mathbb{R}^{2}}\left\vert \sup_{\alpha \in
\lbrack \mathcal{A}](\frac{m-1}{n},\frac{m}{n})}\!\!E_{P}\left[ \int_{\frac{%
m-1}{n}}^{\frac{m}{n}}\partial _{x}H_{\frac{m}{n}}\left( x+\int_{\frac{m-1}{n%
}}^{s}a_{s}^{(1)}ds,y+\int_{\frac{m-1}{n}}^{s}a_{s}^{(2)}dB_{s}^{(2)}\right)
a_{s}^{(1)}ds\right. \right. \\
& \hspace{3cm}\left. +\frac{1}{2}\int_{\frac{m-1}{n}}^{\frac{m}{n}}\partial
_{yy}^{2}H_{\frac{m}{n}}\left( x+\int_{\frac{m-1}{n}}^{s}a_{s}^{(1)}ds,y+%
\int_{\frac{m-1}{n}}^{s}a_{s}^{(2)}dB_{s}^{(2)}\right) (a_{s}^{(2)})^{2}ds%
\right] \\
& \hspace{6cm}-\frac{1}{n}G\left( {\partial _{x}H_{\frac{m}{n}}\left(
x,y\right) },{\partial _{yy}^{2}H_{\frac{m}{n}}\left( x,y\right) }\right) %
\Bigg\vert \\
\leq & \frac{C}{n}\sum_{m=1}^{n}\sup_{z\in \mathbb{R}^{2}}\left\vert
\sup_{\alpha \in \lbrack \mathcal{A}](\frac{m-1}{n},\frac{m}{n})}\!\!E_{P}%
\left[ \sup_{s\in \lbrack \frac{m-1}{n},\frac{m}{n}]}\left( \left\vert \int_{%
\frac{m-1}{n}}^{s}a_{s}^{(1)}ds\right\vert +\left\vert \int_{\frac{m-1}{n}%
}^{s}a_{s}^{(2)}dB_{s}^{(2)}\right\vert \right) \right. \right. \\
& \hspace{4cm}+\left. \left. \sup_{s\in \lbrack \frac{m-1}{n},\frac{m}{n}%
]}\left( \left\vert \int_{\frac{m-1}{n}}^{s}a_{s}^{(1)}ds\right\vert ^{\beta
}+\left\vert \int_{\frac{m-1}{n}}^{s}a_{s}^{(2)}dB_{s}^{(2)}\right\vert
^{\beta }\right) \right] \right\vert \\
& \rightarrow 0\ \text{ as }n\rightarrow \infty ,
\end{align*}%
where $C$ is a constant that depends only on $\overline{\mu },\underline{\mu 
},\overline{\sigma }^{2}$, the uniform bound of $\partial
_{xx}^{2}H_{t},\partial _{xy}^{2}H_{t}$, and constant $L$ in (2).

\medskip

\noindent Prove (5): Use Ito's formula to check that 
\begin{align*}
& \sup_{(x,y)\in \mathbb{R}^{2}}|H_{1}(x,y)-u(x,y)| \\
=& \sup_{(x,y)\in \mathbb{R}^{2}}\left\vert \sup_{a\in \lbrack \mathcal{A}%
](1,1+h)}E_{P}\left[ \int_{1}^{1+h}\partial _{x}u\left(
x+\int_{1}^{s}a_{s}^{(1)}ds,y+\int_{1}^{s}a_{s}^{(2)}dB_{s}^{(2)}\right)
a_{s}^{(1)}ds\right. \right. \\
& \left. \left. \hspace{3cm}+\frac{1}{2}\int_{1}^{1+h}\partial
_{yy}^{2}u\left(
x+\int_{1}^{s}a_{s}^{(1)}ds,y+\int_{1}^{s}a_{s}^{(2)}dB_{s}^{(2)}\right)
(a_{s}^{(2)})^{2}ds\right] \right\vert \\
\leq & ~C_{0}h,
\end{align*}%
where the constant $C_{0}$ depends only on $\overline{\mu },\underline{\mu },%
\overline{\sigma }^{2}$ and the uniform bound of $\partial _{x}u,\partial
_{yy}^{2}u$.

Similarly, we can prove that $\sup_{(x,y)\in \mathbb{R}^{2}}|H_{0}(x,y)-\psi
(x,y)|\leq C_{0}h.$\hfill $\blacksquare $

%

\begin{lemma}
\label{lemma-taylor-G}Take $G$ to be the function defined in (\ref%
{G-function}), let $\{H_{t}\}_{t\in \lbrack 0,1]}$ be the functions defined
in (\ref{Ht}), and define $\{L_{m,n}\}_{m=1}^{n}$ by\footnote{%
Again, $z=\left( z_{1},z_{2}\right) =\left( x,y\right) $.} 
\begin{equation}
L_{m,n}(z)=H_{\frac{m}{n}}(z)+\frac{1}{n}G\left( {\partial _{z_{1}}H_{\frac{m%
}{n}}(z)},{\partial _{z_{2}z_{2}}^{2}H_{\frac{m}{n}}(z)}\right) ,\ z\in 
\mathbb{R}^{2}\text{.}  \label{Lmn}
\end{equation}%
For any $\theta \in \Theta $ and $n\geq 1$, define 
\begin{equation*}
S_{n}^{\theta }=\sum_{i=1}^{n}Z_{i}^{\theta },\quad \overline{S}_{n}^{\theta
}=\sum_{i=1}^{n}\overline{Z}_{i}^{\theta },\quad \overline{Z}_{n}^{\theta
}=Z_{n}^{\theta }-E_{P}[Z_{n}^{\theta }|\mathcal{H}_{n-1}^{\theta }].
\end{equation*}%
Then 
\begin{equation}
\lim\limits_{n\rightarrow \infty }\sum_{m=1}^{n}\left\vert \sup_{\theta \in
\Theta }E_{P}\left[ H_{\frac{m}{n}}\left( \frac{S_{m}^{\theta }}{n},\frac{%
\overline{S}_{m}^{\theta }}{\sqrt{n}}\right) \right] -\sup_{\theta \in
\Theta }E_{P}\left[ L_{m,n}\left( \frac{S_{m-1}^{\theta }}{n},\frac{%
\overline{S}_{m-1}^{\theta }}{\sqrt{n}}\right) \right] \right\vert =0.
\label{lemma-taylor-eq1-2d}
\end{equation}%
%
%
%
%
%
%
%
%
%
%
%
%
%
%
%
%
%
%
%
%
%
%
%
%
%
%
%
%
%
%
%
\end{lemma}

\noindent \textbf{Proof:} We need only prove 
\begin{align}
\lim\limits_{n\rightarrow \infty }\sum_{m=1}^{n}\left\vert \sup_{\theta \in
\Theta }E_{P}\left[ H_{\frac{m}{n}}\left( \frac{S_{m}^{\theta }}{n},\frac{%
\overline{S}_{m}^{\theta }}{\sqrt{n}}\right) \right] -e(m,n)\right\vert & =0%
\text{ ~and}  \label{var-taylor1-2d} \\
\lim\limits_{n\rightarrow \infty }\sum_{m=1}^{n}\left\vert
e(m,n)-\sup_{\theta \in \Theta }E_{P}\left[ L_{m,n}\left( \frac{%
S_{m-1}^{\theta }}{n},\frac{\overline{S}_{m-1}^{\theta }}{\sqrt{n}}\right) %
\right] \right\vert & =0,  \label{var-taylor2-2d}
\end{align}%
where $e(m,n)$ is given by 
\begin{align*}
e(m,n)=& \sup_{\theta \in \Theta }E_{P}\left[ H_{\frac{m}{n}}\left( \frac{%
S_{m-1}^{\theta }}{n},\frac{\overline{S}_{m-1}^{\theta }}{\sqrt{n}}\right)
+\partial _{z_{1}}H_{\frac{m}{n}}\left( \frac{S_{m-1}^{\theta }}{n},\frac{%
\overline{S}_{m-1}^{\theta }}{\sqrt{n}}\right) \frac{Z_{m}^{\theta }}{n}%
\right. \\
& \left. +\partial _{z_{2}}H_{\frac{m}{n}}\left( \frac{S_{m-1}^{\theta }}{n},%
\frac{\overline{S}_{m-1}^{\theta }}{\sqrt{n}}\right) \frac{\overline{Z}%
_{m}^{\theta }}{\sqrt{n}}+\partial _{z_{2}z_{2}}^{2}H_{\frac{m}{n}}\left( 
\frac{S_{m-1}^{\theta }}{n},\frac{\overline{S}_{m-1}^{\theta }}{\sqrt{n}}%
\right) \frac{(\overline{Z}_{m}^{\theta })^{2}}{2n}\right] .
\end{align*}

By Lemma~\ref{lemma-ddp-G}, parts (1) and (2), $\exists C>0$, $\beta \in
(0,1)$ such that 
\begin{equation*}
\sup\limits_{t\in \lbrack 0,1]}\sup\limits_{z\in \mathbb{R}^{2}}|\partial
_{z_{i}z_{j}}^{2}H_{t}(z)|\leq C,
\end{equation*}%
\begin{equation*}
\sup\limits_{t\in \lbrack 0,1]}\sup\limits_{{z,z^{\prime }\in \mathbb{R}^{2},%
}{z\neq z}^{\prime }}\frac{|\partial _{z_{i}z_{j}}^{2}H_{t}(z)-\partial
_{z_{i}z_{j}}^{2}H_{t}(z^{\prime })|}{|z-z^{\prime }|^{\beta }}\leq C,\
i,j=1,2.
\end{equation*}%
%
%
%
%
%
%
%
%
%
%
%
%
%
%
%
%
%
%
%
%
%
%
%
%
%
%
%
%
It follows from Taylor's expansion that $\forall \epsilon >0$ $\exists
\delta >0$ (depending only on $C$ and $\epsilon $), such that $\forall
z,z^{\prime }\in \mathbb{R}^{2}$, and $\forall t\in \lbrack 0,1]$,\footnote{%
Here $D_{z}:=(\partial _{z_{i}})_{i=1}^{2}$ and $D_{z}^{2}:=(\partial
_{z_{i}z_{j}}^{2})_{i,j=1}^{2}$.} 
\begin{align}
& \left\vert H_{t}(z+z^{\prime })-H_{t}(z)-D_{z}H_{t}(z)z^{\prime }-\tfrac{1%
}{2}tr\left( z^{\prime }{}^{\top }D_{z}^{2}H_{t}(z)z^{\prime }\right)
\right\vert  \notag \\
\leq & \epsilon |z^{\prime }|^{2}I_{\{|z^{\prime }|<\delta \}}+2C|z^{\prime
}|^{2}I_{\{|z^{\prime }|\geq \delta \}}\text{.}  \label{var-ty-G-2d}
\end{align}%
Set $z=\left( \frac{S_{m-1}^{\theta }}{n},\frac{\overline{S}_{m-1}^{\theta }%
}{\sqrt{n}}\right) $ and $z^{\prime }=\left( \frac{Z_{m}^{\theta }}{n},\frac{%
\overline{Z}_{m}^{\theta }}{\sqrt{n}}\right) $. Use (\ref{var-ty-G-2d}) to
obtain 
\begin{align*}
& \sum_{m=1}^{n}\left\vert \sup_{\theta \in \Theta }E_{P}\left[ H_{\frac{m}{n%
}}\left( \frac{S_{m}^{\theta }}{n},\frac{\overline{S}_{m}^{\theta }}{\sqrt{n}%
}\right) \right] -e(m,n)\right\vert \\
\leq & \frac{C}{2}\sum_{m=1}^{n}\sup_{\theta \in \Theta }E_{P}\left[
\left\vert \frac{Z_{m}^{\theta }}{n}\right\vert ^{2}+\left\vert \frac{%
Z_{m}^{\theta }}{n}\right\vert \left\vert \frac{\overline{Z}_{m}^{\theta }}{%
\sqrt{n}}\right\vert \right] \\
& +\epsilon \sum_{m=1}^{n}\sup_{\theta \in \Theta }E_{P}\left[ \left(
\left\vert \frac{Z_{m}^{\theta }}{n}\right\vert ^{2}+\left\vert \frac{%
\overline{Z}_{m}^{\theta }}{\sqrt{n}}\right\vert ^{2}\right) I_{\left\{ 
\sqrt{\left\vert \frac{Z_{m}^{\theta }}{n}\right\vert ^{2}+\left\vert \frac{%
\overline{Z}_{m}^{\theta }}{\sqrt{n}}\right\vert ^{2}}<\delta \right\} }%
\right] \\
& +2C\sum_{m=1}^{n}\sup_{\theta \in \Theta }E_{P}\left[ \left( \left\vert 
\frac{Z_{m}^{\theta }}{n}\right\vert ^{2}+\left\vert \frac{\overline{Z}%
_{m}^{\theta }}{\sqrt{n}}\right\vert ^{2}\right) I_{\left\{ \sqrt{\left\vert 
\frac{Z_{m}^{\theta }}{n}\right\vert ^{2}+\left\vert \frac{\overline{Z}%
_{m}^{\theta }}{\sqrt{n}}\right\vert ^{2}}\geq \delta \right\} }\right] 
\text{,}
\end{align*}%
and the latter expression converges to $0$ as $n\rightarrow \infty $ and $%
\epsilon \rightarrow 0$. (Convergence is due to the finiteness of $%
\underline{\mu },\overline{\mu }$ and $\overline{\sigma }$.) This proves (%
\ref{var-taylor1-2d}).

Combine with Lemma \ref{tab-proper} and show that $e(m,n)=$ 
\begin{align*}
& \sup_{\theta \in \Theta }E_{P}\left[ H_{\frac{m}{n}}\left( \frac{%
S_{m-1}^{\theta }}{n},\frac{\overline{S}_{m-1}^{\theta }}{\sqrt{n}}\right)
+\partial _{z_{1}}H_{\frac{m}{n}}\left( \frac{S_{m-1}^{\theta }}{n},\frac{%
\overline{S}_{m-1}^{\theta }}{\sqrt{n}}\right) \frac{Z_{m}^{\theta }}{n}%
\right. \\
& \left. \hspace{5cm}+\partial _{z_{2}z_{2}}^{2}H_{\frac{m}{n}}\left( \frac{%
S_{m-1}^{\theta }}{n},\frac{\overline{S}_{m-1}^{\theta }}{\sqrt{n}}\right) 
\frac{(\overline{Z}_{m}^{\theta })^{2}}{2n}\right] \\
=& \sup_{\theta \in \Theta }E_{P}\left[ H_{\frac{m}{n}}\left( \frac{%
S_{m-1}^{\theta }}{n},\frac{\overline{S}_{m-1}^{\theta }}{\sqrt{n}}\right)
+\max_{1\leq k\leq K}E_{P}\Bigg[\partial _{z_{1}}H_{\frac{m}{n}}\left( \frac{%
S_{m-1}^{\theta }}{n},\frac{\overline{S}_{m-1}^{\theta }}{\sqrt{n}}\right) 
\frac{\mu _{k}}{n}\right. \\
& \left. \hspace{6cm}+\partial _{z_{2}z_{2}}^{2}H_{\frac{m}{n}}\left( \frac{%
S_{m-1}^{\theta }}{n},\frac{\overline{S}_{m-1}^{\theta }}{\sqrt{n}}\right) 
\frac{\sigma _{k}^{2}}{2n}\Bigg]\right] \\
=& \sup_{\theta \in \Theta }E_{P}\left[ L_{m,n}\left( \frac{S_{m-1}^{\theta }%
}{n},\frac{\overline{S}_{m-1}^{\theta }}{\sqrt{n}}\right) \right] .
\end{align*}%
{\normalsize \ }This proves (\ref{var-taylor2-2d}), and completes the proof
of (\ref{lemma-taylor-eq1-2d}). 
\hfill {\normalsize \hfill $\blacksquare $\hfill }

{\normalsize \bigskip }

{\normalsize \noindent }\noindent

\noindent \textbf{Proof of Proposition~\ref{propn-CLT}:} We prove it for%
{\normalsize \ $u\in C_{b}^{\infty }(\mathbb{R}^{2})$}. This suffices
because any {\normalsize $u\in C_{b,Lip}(\mathbb{R}^{2})$ }can be
approximated uniformly by a sequence of functions in $C_{b}^{\infty }$%
{\normalsize $(\mathbb{R}^{2})$ }(see Approximation Lemma in Feller (1971,
Ch. VIII)). The proof also assumes \underline{$\sigma $}$>0$.

For small enough $h>0$, we continue to use $\{H_{t}(x,y)\}_{t\in \lbrack
0,1+h]}$ as defined in (\ref{Ht}). Let $\{L_{m,n}(x,y)\}_{m=1}^{n}$ be the
functions defined in (\ref{Lmn}). By direct calculation we obtain%
{\normalsize \ 
\begin{align*}
& \quad \sup_{\theta \in \Theta }E_{P}\left[ H_{1}\left( \frac{S_{n}^{\theta
}}{n},\frac{\overline{S}_{n}^{\theta }}{\sqrt{n}}\right) \right] -H_{0}(0,0)
\\
& =\sum\limits_{m=1}^{n}\left\{ \sup_{\theta \in \Theta }E_{P}\left[ H_{%
\frac{m}{n}}\left( \frac{S_{m}^{\theta }}{n},\frac{\overline{S}_{m}^{\theta }%
}{\sqrt{n}}\right) \right] -\sup_{\theta \in \Theta }E_{P}\left[ H_{\frac{m-1%
}{n}}\left( \frac{S_{m-1}^{\theta }}{n},\frac{\overline{S}_{m-1}^{\theta }}{%
\sqrt{n}}\right) \right] \right\} \\
& =\sum\limits_{m=1}^{n}\left\{ \sup_{\theta \in \Theta }E_{P}\left[ H_{%
\frac{m}{n}}\left( \frac{S_{m}^{\theta }}{n},\frac{\overline{S}_{m}^{\theta }%
}{\sqrt{n}}\right) \right] -\sup_{\theta \in \Theta }E_{P}\left[
L_{m,n}\left( \frac{S_{m-1}^{\theta }}{n},\frac{\overline{S}_{m-1}^{\theta }%
}{\sqrt{n}}\right) \right] \right\} \\
& \quad \ +\sum_{m=1}^{n}\left\{ \sup_{\theta \in \Theta }E_{P}\left[
L_{m,n}\left( \frac{S_{m-1}^{\theta }}{n},\frac{\overline{S}_{m-1}^{\theta }%
}{\sqrt{n}}\right) \right] -\sup_{\theta \in \Theta }E_{P}\left[ H_{\frac{m-1%
}{n}}\left( \frac{S_{m-1}^{\theta }}{n},\frac{\overline{S}_{m-1}^{\theta }}{%
\sqrt{n}}\right) \right] \right\} \\
& =:I_{1n}+I_{2n}\text{.}
\end{align*}%
}%
Application of Lemma~\ref{lemma-taylor-G} implies that $|I_{1n}|\rightarrow
0 $ as $n\rightarrow \infty .$ 
Lemma~\ref{lemma-ddp-G} implies 
\begin{align*}
|I_{2n}|& \leq \sum_{m=1}^{n}\sup_{\theta \in \Theta }E_{P}\left[ \left\vert
L_{m,n}\left( \frac{S_{m-1}^{\theta }}{n},\frac{\overline{S}_{m-1}^{\theta }%
}{\sqrt{n}}\right) -H_{\frac{m-1}{n}}\left( \frac{S_{m-1}^{\theta }}{n},%
\frac{\overline{S}_{m-1}^{\theta }}{\sqrt{n}}\right) \right\vert \right] \\
& \leq \sum_{m=1}^{n}\sup\limits_{(x,y)\in \mathbb{R}^{2}}\left\vert
L_{m,n}(x,y)-H_{\frac{m-1}{n}}(x,y)\right\vert \\
& \rightarrow 0\text{ ~as }n\rightarrow \infty ,
\end{align*}%
which implies that 
\begin{equation*}
\lim_{n\rightarrow \infty }\left\vert \sup_{\theta \in \Theta }E_{P}\left[
H_{1}\left( \frac{S_{n}^{\theta }}{n},\frac{\overline{S}_{n}^{\theta }}{%
\sqrt{n}}\right) \right] -H_{0}(0,0)\right\vert =0\text{.}
\end{equation*}%
Combine the latter with Lemma~\ref{lemma-ddp-G}, part (5), to obtain 
\begin{align*}
& \left\vert V-\sup_{a\in \lbrack \mathcal{A}](0,1)}E_{P}\left[ u\left(
\int_{0}^{1}a_{s}^{(1)}ds,\int_{0}^{1}a_{s}^{(2)}dB_{s}^{(2)}\right) \right]
\right\vert \\
=& \lim_{n\rightarrow \infty }\left\vert \sup_{\theta \in \Theta }E_{P}\left[
u\left( \frac{S_{n}^{\theta }}{n},\frac{\overline{S}_{n}^{\theta }}{\sqrt{n}}%
\right) \right] -\sup_{a\in \lbrack \mathcal{A}](0,1)}E_{P}\left[ u\left(
\int_{0}^{1}a_{s}^{(1)}ds,\int_{0}^{1}a_{s}^{(2)}dB_{s}^{(2)}\right) \right]
\right\vert \\
\leq & \lim_{n\rightarrow \infty }\left\vert \sup_{\theta \in \Theta }E_{P} 
\left[ \varphi \left( \frac{S_{n}^{\theta }}{n},\frac{\overline{S}%
_{n}^{\theta }}{\sqrt{n}}\right) \right] -\sup_{\theta \in \Theta }E_{P}%
\left[ H_{1}\left( \frac{S_{n}^{\theta }}{n},\frac{\overline{S}_{n}^{\theta }%
}{\sqrt{n}}\right) \right] \right\vert \\
& +\lim_{n\rightarrow \infty }\left\vert \sup_{\theta \in \Theta }E_{P}\left[
H_{1}\left( \frac{S_{n}^{\theta }}{n},\frac{\overline{S}_{n}^{\theta }}{%
\sqrt{n}}\right) \right] -H_{0}(0,0)\right\vert \\
& +\left\vert H_{0}(0,0)-\sup_{a\in \lbrack \mathcal{A}](0,1)}E_{P}\left[
u\left( \int_{0}^{1}a_{s}^{(1)}ds,\int_{0}^{1}a_{s}^{(2)}dB_{s}^{(2)}\right) %
\right] \right\vert \leq C_{0}h\text{,}
\end{align*}%
where the constant $C_{0}$ depends only on $\underline{\mu },\overline{\mu },%
\overline{\sigma }$ and the uniform bound of $\partial _{x}u$ and $\partial
_{yy}^{2}u$. By the arbitrariness of $h$, the proof of (\ref{lemma-CLT-eq})
is completed.

Finally, prove (\ref{lemma-CLT-eq-extreme}). Let $G$ be defined by (\ref%
{G-function}), and define, for all $(p,q)\in ${\normalsize $\mathbb{R}^{2}$}%
, 
\begin{equation*}
G^{ext}(p,q)=\sup_{(\mu ,\sigma ^{2})\in \mathcal{A}^{ext}}\left[ \mu p+%
\frac{1}{2}\sigma ^{2}q\right] \text{.}
\end{equation*}%
Then%
\begin{equation}
G(p,q)=G^{ext}(p,q)\quad \forall (p,q)\in \mathbb{R}^{2}.  \label{compare-G}
\end{equation}%
The proof is completed by applying a Comparison Theorem (Peng (2019, Theorem
C.2.5)). {\normalsize \hfill $\blacksquare $}

{\normalsize \bigskip }

{\normalsize \noindent }\noindent

\noindent \textbf{Proof of Theorem~\ref{thm-bandits}}:{\normalsize \ }All
the results can be obtained from Proposition \ref{propn-CLT}.%
\ That $u$ need only satisfy continuity and the stated growth condition is
implied by Lemma 2.4.12 and Exercise 2.5.7 in Peng (2019) (or by Rosenthal's
inequality in Zhang (2016)). For the convenience of readers, we provide a
proof in the Online Appendix.{\normalsize \hfill $\blacksquare $}

\subsection{Proof of Theorem \protect\ref{thm-strategies}}

We are given that {\normalsize $u(x,y)$ }is increasing in $x$ and concave in 
$y$, and $(\overline{\mu },\underline{\sigma }^{2})\in ${\normalsize $%
\mathcal{A} $. }

For any $t\in \lbrack 0,1]$ and {\normalsize $(x,y)\in \mathbb{R}^{2}$},
define the function{\normalsize \ 
\begin{equation*}
v(t,x,y)=E_{P}[u(x+(1-t)\overline{\mu },y+\underline{\sigma }%
(B_{1}^{(2)}-B_{t}^{(2)}))].
\end{equation*}%
}Then{\normalsize \ 
\begin{equation*}
v(0,0,0)=E_{P}[u(\overline{\mu },\underline{\sigma }B_{1}^{(2)}]=\int u(%
\overline{\mu },\cdot )d\mathbb{N}(0,\underline{\sigma }^{2}).
\end{equation*}%
}By the (classic) Feynman-Kac formula (Mao (2008, Theorem 2.8.3)), $v$ is
the solution of the (linear parabolic) PDE\ 
\begin{equation}
\left\{ 
\begin{array}{l}
\partial _{t}v(t,x,y)+\overline{\mu }\partial _{x}v(t,x,y)+\frac{1}{2}%
\underline{\sigma }^{2}\partial _{yy}^{2}v(t,x,y)=0\text{, }\left(
t,x,y\right) \in \lbrack 0,1)\times \mathbb{R}^{2} \\ 
v(1,x,y)=u(x,y).%
\end{array}%
\right.  \label{pde-2}
\end{equation}%
Since $u(x,y)$ is increasing in $x$ and concave in $y$, it follows that $%
v(t,x,y)$ is increasing in $x$ and concave in $y$ for any $t\in \lbrack 0,1]$%
, that is, \ 
\begin{equation*}
\partial _{x}v(t,x,y)\geq 0\ \text{ and }\ \partial _{yy}^{2}v(t,x,y)\leq 0%
\text{, \ \ }\forall (t,x,y)\in \lbrack 0,1)\times \mathbb{R}^{2}\text{.}
\end{equation*}%
Given also $(\overline{\mu },\underline{\sigma }^{2})\in ${\normalsize $%
\mathcal{A}$}, it follows that{\normalsize 
\begin{equation*}
\sup_{(\mu ,\sigma ^{2})\in \mathcal{A}}\left\{ \mu \partial _{x}v+\tfrac{1}{%
2}\sigma ^{2}\partial _{yy}^{2}v\right\} =\overline{\mu }\partial _{x}v+%
\tfrac{1}{2}\underline{\sigma }^{2}\partial _{yy}^{2}v,
\end{equation*}%
}and hence that $v$ solves the PDE (\ref{pde-1}). By uniqueness of the
solution (Lemma \ref{lemma-pde}), and (\ref{Vv}), conclude that{\normalsize %
\ 
\begin{equation*}
V=v(0,0,0)=\int u(\overline{\mu },\cdot )d\mathbb{N}(0,\underline{\sigma }%
^{2}).\text{ \ \ \ }\blacksquare 
\end{equation*}%
\hfill \hfill\ \ \medskip }

\subsection{Proof of Theorem \protect\ref{thm-tradeoff}}

\noindent Throughout we assume that $\mathcal{A}=\{(\mu _{1},\sigma
_{1}^{2}),(\mu _{2},\sigma _{2}^{2})\}$. \quad \smallskip \newline
\textbf{Proof of (i)}: \ The proof consists of three steps.\newline
\textbf{Step 1:} From Theorem \ref{thm-bandits}(i) and (\ref{Vv}), it
follows that%
\begin{equation*}
\lim_{n\rightarrow \infty }V_{n}=\lim_{n\rightarrow \infty }\sup_{\theta \in
\Theta }E_{P}\left[ u\left( \frac{S_{n}^{\theta }}{n},\frac{\overline{S}%
_{n}^{\theta }}{\sqrt{n}}\right) \right] =v(0,0,0)
\end{equation*}%
where $v(t,x,y)$ solves the PDE (\ref{pde-1}).{\scriptsize \ }

\noindent \textbf{Step 2:} Prove that the following function $v$ solves the
above PDE: 
\begin{align}
\hat{v}(t,x,y)=& E_{P}[u(x+(1-t)\mu _{1},y+\sigma
_{1}(B_{1}^{(2)}-B_{t}^{(2)}))]  \label{wt} \\
=& \int_{\mathbb{R}}u(x+(1-t)\mu _{1},y+\sqrt{1-t}\sigma _{1}r)\frac{1}{%
\sqrt{2\pi }}e^{-\frac{r^{2}}{2}}dr  \notag
\end{align}%
By the Feynman-Kac formula, $\hat{v}$ solves 
\begin{equation}
\left\{ 
\begin{array}{l}
\partial _{t}\hat{v}(t,x,y)+\mu _{1}\partial _{x}\hat{v}(t,x,y)+\frac{1}{2}%
\sigma _{1}^{2}\partial _{yy}^{2}\hat{v}(t,x,y)=0\text{, }\left(
t,x,y\right) \in \lbrack 0,1)\times \mathbb{R}^{2} \\ 
\hat{v}(1,x,y)=u(x,y).%
\end{array}%
\right.  \label{pde-3}
\end{equation}

\noindent From (\ref{wt}) and assumption (\ref{AAi}), it follows that, for
all $\left( t,x,y\right) \in \lbrack 0,1)\times \mathbb{R}^{2}$, 
\begin{equation*}
\tfrac{1}{2}\sigma _{1}^{2}\partial _{yy}^{2}\hat{v}(t,x,y)+\mu _{1}\partial
_{x}\hat{v}(t,x,y)\geq \tfrac{1}{2}\sigma _{2}^{2}\partial _{yy}^{2}\hat{v}%
(t,x,y)+\mu _{2}\partial _{x}\hat{v}(t,x,y)\text{,}
\end{equation*}%
that is, 
\begin{equation}
\sup\limits_{(\mu ,\sigma ^{2})\in \mathcal{A}}\left\{ \mu \partial _{x}\hat{%
v}+\frac{1}{2}\sigma ^{2}\partial _{yy}^{2}\hat{v}\right\} =\mu _{1}\partial
_{x}\hat{v}+\tfrac{1}{2}\sigma _{1}^{2}\partial _{yy}^{2}\hat{v}\text{.}
\label{pdei}
\end{equation}%
Thus $\hat{v}$ solves the PDE (\ref{pde-1}). By uniqueness of the solution
(Lemma \ref{lemma-pde}), conclude that 
\begin{equation*}
\lim_{n\rightarrow \infty }V_{n}=v(0,0,0)=\hat{v}(0,0,0)=\int u(\mu
_{1},\cdot )d\mathbb{N}(0,\sigma _{1}^{2}).
\end{equation*}%

\noindent \textbf{Step 3:} If $\theta ^{\ast }$ denotes the strategy of
choosing arm 1 always, then, using Step 1, 
\begin{equation*}
\lim_{n\rightarrow \infty }E_{P}\left[ u\left( \frac{S_{n}^{\theta ^{\ast }}%
}{n},\frac{\overline{S}_{n}^{\theta ^{\ast }}}{\sqrt{n}}\right) \right]
=E_{P}[u(\mu _{1},\sigma _{1}B_{1}^{(2)})]=v\left( 0,0,0\right) =V\text{.}
\end{equation*}%
Hence $\theta ^{\ast }$ is asymptotically optimal.

\medskip

\noindent \textbf{Proof of (iii)}: Case 1 ($\alpha \leq \frac{\mu _{1}-\mu
_{2}}{\sigma _{1}^{2}-\sigma _{2}^{2}}$): {Define $v$} by (\ref{wt}). {%
Although $u$ is not twice differentiable, we can calculate $\partial _{x}v$
and $\partial _{yy}^{2}v$ directly to obtain $\partial _{x}v=1$ and $%
\partial _{yy}^{2}v=-2\alpha \Phi (\frac{-y}{\sigma _{1}\sqrt{1-t}})$. }%
Therefore,%
\begin{eqnarray*}
\alpha &<&\frac{\mu _{1}-\mu _{2}}{\sigma _{1}^{2}-\sigma _{2}^{2}}%
\Longrightarrow \\
\mu _{1}-\alpha \Phi (\frac{-y}{\sqrt{1-t}\sigma _{1}})\sigma _{1}^{2}
&>&\mu _{2}-\alpha \Phi (\frac{-y}{\sqrt{1-t}\sigma _{1}})\sigma
_{2}^{2}\Longrightarrow \\
\mu _{1}\partial _{x}v+\tfrac{1}{2}\sigma _{1}^{2}\partial _{yy}^{2}v &>&\mu
_{2}\partial _{x}v+\tfrac{1}{2}\sigma _{2}^{2}\partial _{yy}^{2}v.
\end{eqnarray*}%
Proceed as in the proof of (i).\footnote{%
If we assume the reverse inequality in (\ref{AAii}), then corresponding
implications fail. For example, if $y>0$ is sufficiently large which would
make $\Phi (\frac{-y}{\sqrt{1-t}\sigma })$ close to zero for $\sigma =\sigma
_{1},\sigma _{2}$. $t\geq 0$, then the last two inequalities above could
remain valid even though $\alpha >\left( \mu _{1}-\mu _{2}\right) /\left(
\sigma _{1}^{2}-\sigma _{2}^{2}\right) $.}

\smallskip

\noindent Case 2 ($\underline{\alpha }<\alpha <\overline{\alpha }$): To
prove that single-arm strategies are not asymptotically optimal, it is
enough to show that 
\begin{equation}
V=\sup_{a\in \lbrack \mathcal{A}](0,1)}E_{P}\left[ u\left(
\int_{0}^{1}a_{s}^{(1)}ds,\int_{0}^{1}a_{s}^{(2)}dB_{s}^{(2)}\right) \right]
>\max_{i=1,2}E_{P}\left[ u\left( \mu _{i},\sigma _{i}B_{1}^{(2)}\right) %
\right] \text{.}  \label{Vdominate}
\end{equation}

Consider the bandit problem with set of arms given by%
\begin{equation*}
\mathcal{A}^{\prime }=\{(\mu _{1},\sigma _{1}^{2}),(\mu _{2},\sigma
_{2}^{2}),(\mu _{3},\sigma _{3}^{2})\}\text{,}
\end{equation*}%
where $(\mu _{3},\sigma _{3}^{2})=(1-\lambda )(\mu _{1},\sigma
_{1}^{2})+\lambda (\mu _{2},\sigma _{2}^{2})$ for some $0<\lambda <1$ to be
selected below. Because $\mathcal{A}^{\prime }$ and $\mathcal{A}$ have the
identical extreme points, Proposition~\ref{propn-CLT} implies that 
\begin{align*}
V=& \sup_{a\in \lbrack \mathcal{A}](0,1)}E_{P}\left[ u\left(
\int_{0}^{1}a_{s}^{(1)}ds,\int_{0}^{1}a_{s}^{(2)}dB_{s}^{(2)}\right) \right]
\\
=& \sup_{a\in \lbrack \mathcal{A}^{\prime }](0,1)}E_{P}\left[ u\left(
\int_{0}^{1}a_{s}^{(1)}ds,\int_{0}^{1}a_{s}^{(2)}dB_{s}^{(2)}\right) \right] 
\text{.}
\end{align*}%
Take 
\begin{equation}
(\hat{a}_{s}^{(1)},\hat{a}_{s}^{(2)})=(\mu _{1},\sigma
_{1})I_{\{W_{s}^{\sigma _{1},\sigma _{3}}\geq 0\}}+(\mu _{3},\sigma
_{3})I_{\{W_{s}^{\sigma _{1},\sigma _{3}}<0\}}\text{,}  \label{ahat}
\end{equation}%
where 
\begin{equation*}
W_{t}^{\sigma _{1},\sigma _{3}}=\int_{0}^{t}\left( \sigma
_{1}I_{\{W_{s}^{\sigma _{1},\sigma _{3}}\geq 0\}}+\sigma
_{3}I_{\{W_{s}^{\sigma _{1},\sigma _{3}}<0\}}\right) dB_{s}^{(2)}\text{;}
\end{equation*}
$W_{s}^{\sigma _{1},\sigma _{3}}$ is an oscillating Brownian motion, that
is, the solution of the stochastic differential equation (SDE) 
\begin{equation*}
W_{t}^{\sigma _{1},\sigma _{3}}=\int_{0}^{t}\left( \sigma
_{1}I_{\{W_{s}^{\sigma _{1},\sigma _{3}}\geq 0\}}+\sigma
_{3}I_{\{W_{s}^{\sigma _{1},\sigma _{3}}<0\}}\right) dB_{s}^{(2)}\text{.}
\end{equation*}%
By Keilson and Wellner (1978, Theorem 1), the probability density of $%
W_{t}^{\sigma _{1},\sigma _{3}}$ is $q\left( t,\cdot \right) $, where%
\begin{equation}
q\left( t,y\right) =\left\{ 
\begin{array}{ccc}
q^{\ast }\left( y;\sigma _{1}^{2}t\right) \left[ \frac{2\sigma _{3}}{\sigma
_{1}+\sigma _{3}}\right] &  & y\geq 0 \\ 
&  &  \\ 
q^{\ast }\left( y;\sigma _{3}^{2}t\right) \left[ \frac{2\sigma _{1}}{\sigma
_{1}+\sigma _{3}}\right] &  & y<0%
\end{array}%
\right.  \label{qdensity}
\end{equation}%
and $q^{\ast }(y;\sigma ^{2})=\frac{1}{\sqrt{2\pi }\sigma }\exp \left( -({%
y/\sigma })^{2}{/2}\right) $ is the pdf for $\mathbb{N}\left( 0,\sigma
^{2}\right) $. Using this pdf, we can calculate 
\begin{align*}
& E_{P}\left[ u\left( \int_{0}^{1}\hat{a}_{s}^{(1)}ds,\int_{0}^{1}\hat{a}%
_{s}^{(2)}dB_{s}^{(2)}\right) \right] \\
=& E_{P}\left[ \int_{0}^{1}\left( \mu _{1}I_{\{W_{s}^{\sigma _{1},\sigma
_{3}}\geq 0\}}+\mu _{2}I_{\{W_{s}^{\sigma _{1},\sigma _{3}}<0\}}\right) ds%
\right] -\alpha E_{P}\left[ \left( W_{1}^{\sigma _{1},\sigma _{3}}\right)
^{2}I_{\{W_{1}^{\sigma _{1},\sigma _{3}}\leq 0\}}\right] \\
=& \mu _{1}\int_{0}^{1}P(W_{s}^{\sigma _{1},\sigma _{3}}\geq 0)ds+\mu
_{3}\int_{0}^{1}P(W_{s}^{\sigma _{1},\sigma _{3}}<0)ds-\alpha \int_{-\infty
}^{0}y^{2}q(1,y)dy \\
=& \mu _{1}\int_{0}^{1}\int_{0}^{\infty }q(s,y)dyds+\mu
_{3}\int_{0}^{1}\int_{-\infty }^{0}q(s,y)dyds-\alpha \int_{-\infty
}^{0}y^{2}q(1,y)dy \\
=& \mu _{1}\frac{\sigma _{3}}{\sigma _{1}+\sigma _{3}}+\mu _{3}\frac{\sigma
_{1}}{\sigma _{1}+\sigma _{3}}-\alpha \frac{\sigma _{1}\sigma _{3}^{2}}{%
\sigma _{1}+\sigma _{3}}\text{.}
\end{align*}

Verify the inequality 
\begin{equation*}
\mu _{1}\frac{\sigma _{3}}{\sigma _{1}+\sigma _{3}}+\mu _{3}\frac{\sigma _{1}%
}{\sigma _{1}+\sigma _{3}}-\alpha \frac{\sigma _{1}\sigma _{3}^{2}}{\sigma
_{1}+\sigma _{3}}>\mu _{1}-\alpha \frac{\sigma _{1}^{2}}{2}=E_{P}\left[
u\left( \mu _{1},\sigma _{1}B_{1}^{(2)}\right) \right] \text{,}
\end{equation*}%
and deduce that 
\begin{equation*}
\alpha >\frac{2(\mu _{1}-\mu _{3})}{(\sigma _{1}+2\sigma _{3})(\sigma
_{1}-\sigma _{3})}=\frac{\mu _{1}-\mu _{2}}{\sigma _{1}^{2}-\sigma
_{2}^{2}+f(\lambda )}
\end{equation*}%
where $f(\lambda )=\left( \sigma _{1}\sqrt{(1-\lambda )\sigma
_{1}^{2}+\lambda \sigma _{2}^{2}}-\sigma _{1}^{2}\right) /2\lambda $. It can
be verified that $f^{\prime }(\lambda )<0$ for $\lambda \in (0,1)$ and $%
\lim_{\lambda \rightarrow 0}f(\lambda )=\left( \sigma _{2}^{2}-\sigma
_{1}^{2}\right) /4.$

Therefore, for any $\alpha >\underline{\alpha }=\frac{4(\mu _{1}-\mu _{2})}{%
3(\sigma _{1}^{2}-\sigma _{2}^{2})}$, there exists $\lambda _{0}\in (0,1)$
such that 
\begin{equation*}
\alpha >\frac{\mu _{1}-\mu _{2}}{\sigma _{1}^{2}-\sigma _{2}^{2}+f(\lambda
_{0})}>\frac{4(\mu _{1}-\mu _{2})}{3(\sigma _{1}^{2}-\sigma _{2}^{2})}.
\end{equation*}%
Choose $\lambda =\lambda _{0}$ in the definition (\ref{ahat}) of $\hat{a}=(%
\hat{a}_{s}^{(1)},\hat{a}_{s}^{(2)})$ and deduce that 
\begin{align*}
V=& \sup_{a\in \lbrack \mathcal{A}](0,1)}E_{P}\left[ u\left(
\int_{0}^{1}a_{s}^{(1)}ds,\int_{0}^{1}a_{s}^{(2)}dB_{s}^{(2)}\right) \right]
\\
\geq & E_{P}\left[ u\left( \int_{0}^{1}\hat{a}_{s}^{(1)}ds,\int_{0}^{1}\hat{a%
}_{s}^{(2)}dB_{s}^{(2)}\right) \right] \\
>& E_{P}\left[ u\left( \mu _{1},\sigma _{1}B_{1}^{(2)}\right) \right] .
\end{align*}%
That is, when $\alpha >\underline{\alpha }$, then specializing in arm 1 is
NOT asymptotically optimal.

An analogous argument proves that specializing in arm 2 is not
asymptotically optimal if $\alpha <\overline{\alpha }$. Details are provided
in the Online Appendix.

\medskip

\noindent \textbf{Proof of (iv):\ } It remains to prove only the claim for
the case \underline{$\alpha $}$^{\prime }<\alpha $. The proof is similar to
that for (iii). Specifically, prove that (\ref{Vdominate}) is satisfied for
the process $(\hat{a}_{s}^{(1)},\hat{a}_{s}^{(2)})$ where{\scriptsize \ } 
\begin{equation*}
(\hat{a}_{s}^{(1)},\hat{a}_{s}^{(2)})=(\mu _{1},\sigma
_{1})I_{\{W_{s}^{\sigma _{2},\sigma _{1}}<0\}}+(\mu _{2},\sigma
_{2})I_{\{W_{s}^{\sigma _{2},\sigma _{1}}\geq 0\}}\text{,}
\end{equation*}%
and $W_{s}^{\sigma _{2},\sigma _{1}}$ is the oscillating Brownian motion
given by 
\begin{equation*}
W_{t}^{\sigma _{2},\sigma _{1}}=\int_{0}^{t}\left( \sigma
_{1}I_{\{W_{s}^{\sigma _{2},\sigma _{1}}<0\}}+\sigma _{2}I_{\{W_{s}^{\sigma
_{2},\sigma _{1}}\geq 0\}}\right) dB_{s}^{(2)}\text{.}
\end{equation*}%
The process $W_{t}^{\sigma _{2},\sigma _{1}}$ admits a probability density
analogous to (\ref{qdensity}). \medskip

\noindent \noindent \textbf{Proof of (v):} For $i\geq 1$, we have $%
Z_{i}^{\theta ^{\ast }}=X_{k,i}$ where $\theta _{i}^{\ast }=k$, and $%
\{X_{k,i}:i\geq 1\}$ are i.i.d. Then 
\begin{equation*}
E_{P}\left[ \varphi \left( \frac{1}{n}\sum_{i=1}^{n}Z_{i}^{\theta ^{\ast
}}\right) \right] =E_{P}\left[ \varphi \left( \frac{\psi _{n}}{n}\frac{%
\sum_{i=1}^{\psi _{n}}X_{1,i}}{\psi _{n}}+\frac{n-\psi _{n}}{n}\frac{%
\sum_{i=1}^{n-\psi _{n}}X_{2,i}}{n-\psi _{n}}\right) \right]
\end{equation*}%
Since $\psi _{n}/n\rightarrow \lambda $ as $n\rightarrow \infty $, combine
with the classical LLN for $\{X_{1,i}:i\geq 1\}$ and $\{X_{2,i}:i\geq 1\}$
to obtain 
\begin{equation*}
\lim_{n\rightarrow \infty }E_{P}\left[ \varphi \left( \frac{1}{n}%
\sum_{i=1}^{n}Z_{i}^{\theta ^{\ast }}\right) \right] =\varphi \left( \lambda
\mu _{1}+(1-\lambda )\mu _{2}\right) =\varphi (x^{\ast })\text{.}
\end{equation*}%
Therefore, $\theta ^{\ast }$ is asymptotically optimal because, by
Proposition \ref{propn-CLT}, 
\begin{align*}
V=& \sup_{a\in \lbrack \mathcal{A}](0,1)}E_{P}\left[ u\left(
\int_{0}^{1}a_{s}^{(1)}ds,\int_{0}^{1}a_{s}^{(2)}dB_{s}^{(2)}\right) \right]
\\
=& \sup_{a\in \lbrack \mathcal{A}](0,1)}E_{P}\left[ \varphi \left(
\int_{0}^{1}a_{s}^{(1)}ds\right) \right] \leq \varphi (x^{\ast })\text{.}
\end{align*}%
{\normalsize \hfill }

The remaining assertion is implied by the fact that $\lim_{n\longrightarrow
\infty }U_{n}\left( \theta ^{\mu ,\sigma }\right) =\varphi \left( \mu
\right) $ for each $\left( \mu ,\sigma ^{2}\right) $.%
%
%
%
%
%
%
%
%
%
%
%
%
%
%
%
{\normalsize \hfill $\blacksquare $}

\newpage

\begin{center}
{\LARGE ONLINE APPENDIX }for

"Approximate optimality and the risk/reward tradeoff ... "
\end{center}

\medskip \bigskip

\noindent {\LARGE Lemma:} Our CLT, Proposition \ref{propn-CLT}, is valid
also if $\underline{\sigma }=0$.

\noindent \textbf{Proof:} As in the proof of Proposition \ref{propn-CLT}, it
suffices to take $u\in ${\normalsize $C_{b}^{\infty }(\mathbb{R}^{2})$. }

Given $\underline{\sigma }=0$, we add a perturbation to the random returns
of the $K$ arms. For any $1\leq k\leq K$ and $n\geq 1$, let $%
X_{k,n}^{\epsilon }=X_{k,n}+\epsilon \zeta _{n}$, where $\epsilon >0$ is a
fixed small constant and $\{\zeta _{n}\}$ is a sequence of i.i.d. standard
normal random variables, independent with $\{X_{k,n}\}$. Then, for any $%
\theta \in \Theta $ and $n\geq 1$, the corresponding reward is denoted by $%
Z_{n}^{\theta ,\epsilon }=Z_{n}^{\theta }+\epsilon \zeta _{n}$, and the
corresponding set of mean-variance pairs is denoted by 
\begin{equation*}
\mathcal{A}_{\epsilon }=\{(\mu _{k,\epsilon },\sigma _{k,\epsilon
}^{2}):1\leq k\leq K\},
\end{equation*}%
where $\mu _{k,\epsilon }=\mu _{k}$ and $\sigma _{k,\epsilon }^{2}=\sigma
_{k}^{2}+\epsilon ^{2}$. The corresponding bounds are $\overline{\mu }%
_{\epsilon },\underline{\mu }_{\epsilon },\overline{\sigma }_{\epsilon }^{2}$%
, and $\underline{\sigma }_{\epsilon }^{2}>0$.

Define 
\begin{equation*}
V_{n}^{\epsilon }=\sup_{\theta \in \Theta }E_{P}\left[ u\left( \frac{%
\sum_{i=1}^{n}Z_{i}^{\theta ,\epsilon }}{n},\frac{\sum_{i=1}^{n}(Z_{i}^{%
\theta ,\epsilon }-E_{P}[Z_{i}^{\theta ,\epsilon }|\mathcal{H}_{i-1}^{\theta
}]}{\sqrt{n}}\right) \right]
\end{equation*}%
By Proposition \ref{propn-CLT} for $\{Z_{n}^{\theta ,\epsilon }\}$,{%
\begin{equation}
\lim_{n\rightarrow \infty }V_{n}^{\epsilon }=\sup_{a\in \lbrack \mathcal{A}%
_{\epsilon }](0,1)}E_{P}\left[ u\left(
\int_{0}^{1}a_{s}^{(1)}ds,\int_{0}^{1}a_{s}^{(2)}dB_{s}^{(2)}\right) \right]
=v_{\epsilon }(0,0,0),  \label{lemma-CLT-epsilon-eq}
\end{equation}%
}where $v_{\epsilon }(t,x,y)$ is the solution of PDE (\ref{pde-1}) with
function $G_{\epsilon }$ instead of $G$, 
\begin{equation}
G_{\epsilon }(p,q)=\sup_{(\mu ,\sigma ^{2})\in \mathcal{A}_{\epsilon }}\left[
\mu p+\frac{1}{2}\sigma ^{2}q\right] \text{, \ \ }\left( p,q\right) \in 
\mathbb{R}^{2}\text{.}  \label{function-G-ep}
\end{equation}%
By Yong and Zhou (1999, Propn. 5.10, Ch. 4), $\ \exists C^{\prime }>0$ such
that 
\begin{equation*}
|v_{\epsilon }(t,x,y)-v(t,x,y)|\leq C^{\prime }\sqrt{\epsilon },\quad
\forall (t,x,y)\in \lbrack 0,1)\times \mathbb{R}^{2}.
\end{equation*}%
We also have 
\begin{equation*}
\left\vert V_{n}-V_{n}^{\epsilon }\right\vert ^{2}\leq C\epsilon ^{2}E_{P} 
\left[ \left\vert \frac{\sum_{i=1}^{n}\zeta _{i}}{n}\right\vert
^{2}+\left\vert \frac{\sum_{i=1}^{n}\zeta _{i}}{\sqrt{n}}\right\vert ^{2}%
\right] \leq 2C\epsilon ^{2}\text{,}
\end{equation*}%
{where the constant }$C$ depends only on the bounds of $\partial _{x}u$ and $%
\partial _{y}u$.

Letting as $\epsilon \rightarrow 0$ in (\ref{lemma-CLT-epsilon-eq}), the CLT
(\ref{lemma-CLT-eq}) is proven for $\underline{\sigma }=0$. Similar
arguments show that (\ref{lemma-CLT-eq-extreme}) is also valid.\hfill $%
\blacksquare $

\bigskip

\noindent {\LARGE Lemma:} Our CLT, Proposition \ref{propn-CLT}, is valid
also if $u$ is continuous and, for some $g\geq 1$ and $c>0$, $|u(x,y)|\leq
c(1+||(x.y)||^{g-1})${\small \ }and {$\sup_{1\leq k\leq
K}E_{P}[|X_{k}|^{g}]<\infty $}.

\noindent \textbf{Proof:} We prove that (\ref{lemma-CLT-eq}) remains valid.
Refer to it as "the CLT." \newline
Step 1: Prove the CLT for any $u\in C_{b}(\mathbb{R}^{2})$ with compact
support (constant outside a compact subset of $\mathbb{R}^{2}$). In this
case, $\forall \epsilon >0$ $\exists \hat{u}\in C_{b,Lip}(\mathbb{R}^{2})$
such that $\sup_{z\in \mathbb{R}^{2}}|u(z)-\hat{u}(z)|\leq \frac{\epsilon }{2%
}$. Then 
\begin{align*}
& \left\vert \sup_{\theta \in \Theta }E_{P}\left[ u\left( \frac{%
S_{n}^{\theta }}{n},\frac{\overline{S}_{n}^{\theta }}{\sqrt{n}}\right) %
\right] -\sup_{a\in \lbrack \mathcal{A}](0,1)}E_{P}[u(%
\int_{0}^{1}a_{s}^{(1)}ds,\int_{0}^{1}a_{s}^{(2)}dB_{s}^{(2)})]\right\vert \\
\leq & \epsilon +\left\vert \sup_{\theta \in \Theta }E_{P}\left[ \hat{u}%
\left( \frac{S_{n}^{\theta }}{n},\frac{\overline{S}_{n}^{\theta }}{\sqrt{n}}%
\right) \right] -\sup_{a\in \lbrack \mathcal{A}](0,1)}E_{P}[\hat{u}%
(\int_{0}^{1}a_{s}^{(1)}ds,\int_{0}^{1}a_{s}^{(2)}dB_{s}^{(2)})]\right\vert
\end{align*}%
Therefore, 
\begin{equation*}
\limsup_{n\rightarrow \infty }\left\vert \sup_{\theta \in \Theta }E_{P}\left[
u\left( \frac{S_{n}^{\theta }}{n},\frac{\overline{S}_{n}^{\theta }}{\sqrt{n}}%
\right) \right] -\sup_{a\in \lbrack \mathcal{A}](0,1)}E_{P}[u(%
\int_{0}^{1}a_{s}^{(1)}ds,\int_{0}^{1}a_{s}^{(2)}dB_{s}^{(2)})]\right\vert
\leq \epsilon ,
\end{equation*}%
which proves the CLT since $\epsilon $ is arbitrary.

\noindent Step 2: Let $u\in C${\normalsize $(\mathbb{R}^{2})$} satisfy the
growth condition $|u(z)|\leq c(1+|z|^{g-1})$ for $g\geq 1$. For any $N>0$, $%
\exists u_{1},u_{2}\in ${\normalsize $C(\mathbb{R}^{2})$ }such that $%
u=u_{1}+u_{2}$, where $u_{1}$ has a compact support and $u_{2}(z)=0$ for $%
|z|\leq N$, and $|u_{2}(z)|\leq |u(z)|$ for all $z$. Then{\normalsize 
\begin{equation*}
|u_{2}(z)|\leq \frac{2c(1+|z|^{g})}{N},\quad \forall z\in \mathbb{R}^{2}%
\text{,}
\end{equation*}%
}and 
\begin{align*}
& \left\vert \sup_{\theta \in \Theta }E_{P}\left[ u\left( \frac{%
S_{n}^{\theta }}{n},\frac{\overline{S}_{n}^{\theta }}{\sqrt{n}}\right) %
\right] -\sup_{a\in \lbrack \mathcal{A}](0,1)}E_{P}[u(%
\int_{0}^{1}a_{s}^{(1)}ds,\int_{0}^{1}a_{s}^{(2)}dB_{s}^{(2)})]\right\vert \\
\leq & \left\vert \sup_{\theta \in \Theta }E_{P}\left[ u_{1}\left( \frac{%
S_{n}^{\theta }}{n},\frac{\overline{S}_{n}^{\theta }}{\sqrt{n}}\right) %
\right] -\sup_{a\in \lbrack \mathcal{A}](0,1)}E_{P}[u_{1}(%
\int_{0}^{1}a_{s}^{(1)}ds,\int_{0}^{1}a_{s}^{(2)}dB_{s}^{(2)})]\right\vert \\
& +\sup_{\theta \in \Theta }E_{P}\left[ \left\vert u_{2}\left( \frac{%
S_{n}^{\theta }}{n},\frac{\overline{S}_{n}^{\theta }}{\sqrt{n}}\right)
\right\vert \right] +\sup_{a\in \lbrack \mathcal{A}](0,1)}E_{P}[|u_{2}(%
\int_{0}^{1}a_{s}^{(1)}ds,\int_{0}^{1}a_{s}^{(2)}dB_{s}^{(2)})|] \\
\leq & \left\vert \sup_{\theta \in \Theta }E_{P}\left[ u_{1}\left( \frac{%
S_{n}^{\theta }}{n},\frac{\overline{S}_{n}^{\theta }}{\sqrt{n}}\right) %
\right] -\sup_{a\in \lbrack \mathcal{A}](0,1)}E_{P}[u_{1}(%
\int_{0}^{1}a_{s}^{(1)}ds,\int_{0}^{1}a_{s}^{(2)}dB_{s}^{(2)})]\right\vert \\
& +\frac{2c}{N}\left( 2+\sup_{\theta \in \Theta }E_{P}\left[ \left\vert 
\frac{S_{n}^{\theta }}{n}\right\vert ^{g}+\left\vert \frac{\overline{S}%
_{n}^{\theta }}{\sqrt{n}}\right\vert ^{g}\right] +\sup_{a\in \lbrack 
\mathcal{A}](0,1)}E_{P}\left[ \left\vert
\int_{0}^{1}a_{s}^{(1)}ds\right\vert ^{g}+\left\vert
\int_{0}^{1}a_{s}^{(2)}dB_{s}^{(2)}\right\vert ^{g}\right] \right)
\end{align*}%
By the Burkholder-Davis-Gundy inequality\ {(Mao (2008, Theorem 1.7.3))}, 
\begin{align*}
& \limsup_{n\rightarrow \infty }\left\vert \sup_{\theta \in \Theta }E_{P} 
\left[ u\left( \frac{S_{n}^{\theta }}{n},\frac{\overline{S}_{n}^{\theta }}{%
\sqrt{n}}\right) \right] -\sup_{a\in \lbrack \mathcal{A}](0,1)}E_{P}[u(%
\int_{0}^{1}a_{s}^{(1)}ds,\int_{0}^{1}a_{s}^{(2)}dB_{s}^{(2)})]\right\vert \\
\leq & \frac{2c}{N}\left( 2+\max \{|\overline{\mu }|^{g},|\underline{\mu }%
|^{g}\}+\overline{\sigma }^{g}+\sup_{n}\sup_{\theta \in \Theta }E_{P}\left[
\left\vert \frac{S_{n}^{\theta }}{n}\right\vert ^{g}+\left\vert \frac{%
\overline{S}_{n}^{\theta }}{\sqrt{n}}\right\vert ^{g}\right] \right) .
\end{align*}%
Since $N$ can be arbitrarily large, it suffices to prove 
\begin{equation*}
\sup_{n}\sup_{\theta \in \Theta }E_{P}\left[ \left\vert \frac{S_{n}^{\theta }%
}{n}\right\vert ^{g}+\left\vert \frac{\overline{S}_{n}^{\theta }}{\sqrt{n}}%
\right\vert ^{g}\right] <\infty
\end{equation*}

\noindent Step 3: Prove the preceding inequality. For any $n$, 
\begin{equation*}
\sup_{\theta \in \Theta }E_{P}\left[ \left\vert \frac{S_{n}^{\theta }}{n}%
\right\vert ^{g}\right] \leq \sup_{\theta \in \Theta }E_{P}\left[ \frac{%
n^{g-1}}{n^{g}}\sum_{i=1}^{n}|Z_{i}^{\theta }|^{g}\right] \leq K\sup_{1\leq
k\leq K}E_{P}[|X_{k}|^{g}].
\end{equation*}%
For $1\leq g\leq 2$, 
\begin{align*}
\left( \sup_{\theta \in \Theta }E_{P}\left[ \left\vert \frac{\overline{S}%
_{n}^{\theta }}{\sqrt{n}}\right\vert ^{g}\right] \right) ^{\frac{2}{g}}\leq
& \sup_{\theta \in \Theta }E_{P}\left[ \left( \frac{\overline{S}_{n}^{\theta
}}{\sqrt{n}}\right) ^{2}\right] \\
=& \frac{1}{n}\sup_{\theta \in \Theta }E_{P}\left[ \left( \overline{S}%
_{n-1}^{\theta }\right) ^{2}+2\overline{S}_{n-1}^{\theta }\overline{Z}%
_{n}^{\theta }+(\overline{Z}_{n}^{\theta })^{2}\right] \\
\leq & \frac{1}{n}\sup_{\theta \in \Theta }E_{P}\left[ \left( \overline{S}%
_{n-1}^{\theta }\right) ^{2}+\overline{\sigma }^{2}\right] \leq \overline{%
\sigma }^{2}.
\end{align*}%
For $g>2$, 
\begin{equation*}
|x+y|^{g}\leq
2^{g}g^{2}|x|^{g}+|y|^{g}+gx|y|^{g-1}sgn(y)+2^{g}g^{2}x^{2}|y|^{g-2},\
\forall x,y\in \mathbb{R}.
\end{equation*}%
Let $T_{k}^{\theta }=\max \{\overline{S}_{k}^{\theta },\overline{S}%
_{k}^{\theta }-\overline{S}_{1}^{\theta },\cdots ,\overline{S}_{k}^{\theta }-%
\overline{S}_{k-1}^{\theta }\}$. Then $T_{k}^{\theta }=\overline{Z}%
_{k}^{\theta }+(T_{k-1}^{\theta })^{+}$ and 
\begin{align*}
& \sup_{\theta \in \Theta }E_{P}[|T_{k}^{\theta }|^{g}] \\
\leq & 2^{g}g^{2}\sup_{\theta \in \Theta }E_{P}[|\overline{Z}_{k}^{\theta
}|^{g}]+\sup_{\theta \in \Theta }E_{P}[|(T_{k-1}^{\theta })^{+}|^{g}] \\
& +g\sup_{\theta \in \Theta }E_{P}[\overline{Z}_{k}^{\theta
}|(T_{k-1}^{\theta })^{+}|^{g-1}]+2^{g}g^{2}\sup_{\theta \in \Theta }E_{P}[(%
\overline{Z}_{k}^{\theta })^{2}|(T_{k-1}^{\theta })^{+}|^{g-2}] \\
\leq & 2^{g}g^{2}\sum_{i=1}^{k}\sup_{\theta \in \Theta }E_{P}[|\overline{Z}%
_{i}^{\theta }|^{g}]+2^{g}g^{2}\sum_{i=2}^{k}\sup_{\theta \in \Theta }E_{P}[(%
\overline{Z}_{i}^{\theta })^{2}|(T_{i-1}^{\theta })^{+}|^{g-2}] \\
\leq & 2^{g}g^{2}\sum_{i=1}^{n}\sup_{\theta \in \Theta }E_{P}[|\overline{Z}%
_{i}^{\theta }|^{g}]+2^{g}g^{2}\overline{\sigma }^{2}\sum_{i=1}^{n}\left(
\sup_{\theta \in \Theta }E_{P}[|(T_{i}^{\theta })^{+}|^{g}]\right) ^{\frac{%
g-2}{g}}\text{.}
\end{align*}%
Let $A_{n}=\sup_{k\leq n}\sup_{\theta \in \Theta }E_{P}[|T_{k}^{\theta
}|^{g}]$. Then, by Young's inequality {(Peng (2019, Lemma 1.4.1))},\footnote{%
$\mid ab\mid \leq p^{-1}\mid a\mid ^{p}+q^{-1}\mid a\mid ^{q}$ if $%
1<p,q<\infty $ and $p^{-1}+q^{-1}=1$.}

\begin{align*}
A_{n}\leq & 2^{g}g^{2}\sum_{i=1}^{n}\sup_{\theta \in \Theta }E_{P}[|%
\overline{Z}_{i}^{\theta }|^{g}]+2^{g}g^{2}\overline{\sigma }^{2}nA_{n}^{%
\frac{g-2}{g}} \\
\leq & 2^{g}g^{2}\sum_{i=1}^{n}\sup_{\theta \in \Theta }E_{P}[|\overline{Z}%
_{i}^{\theta }|^{g}]+\frac{2}{g}(2^{g}g^{2}\overline{\sigma }^{2}n)^{\frac{g%
}{2}}+\frac{g-2}{g}A_{n}.
\end{align*}%
Therefore, 
\begin{align*}
A_{n}\leq & C_{g,1}\sum_{i=1}^{n}\sup_{\theta \in \Theta }E_{P}[|\overline{Z}%
_{i}^{\theta }|^{g}]+C_{g,2}n^{\frac{g}{2}} \\
\leq & C_{g,1}\sum_{i=1}^{n}\sup_{\theta \in \Theta }E_{P}[|Z_{i}^{\theta
}|^{g}+\max \{|\overline{\mu }|^{g},|\underline{\mu }|^{g}\}]+C_{g,2}n^{%
\frac{g}{2}} \\
\leq & C_{g,1}nK\sup_{1\leq k\leq K}E_{P}[|X_{k}|^{g}]+C_{g,1}n\max \{|%
\overline{\mu }|^{g},|\underline{\mu }|^{g}\}+C_{g,2}n^{\frac{g}{2}}.
\end{align*}%
Finally, 
\begin{align*}
& \sup_{\theta \in \Theta }E_{P}\left[ \left\vert \frac{\overline{S}%
_{n}^{\theta }}{\sqrt{n}}\right\vert ^{g}\right] \leq n^{-\frac{g}{2}}A_{n}
\\
\leq & C_{g,1}n^{1-\frac{g}{2}}K\sup_{1\leq k\leq
K}E_{P}[|X_{k}|^{g}]+C_{g,1}n^{1-\frac{g}{2}}\max \{|\overline{\mu }|^{g},|%
\underline{\mu }|^{g}\}+C_{g,2}.
\end{align*}

\noindent Since $\sup_{1\leq k\leq K}E_{P}[|X_{k}|^{g}]<\infty $, Step 3 is
complete and the Lemma is proven. \hfill $\blacksquare $

\bigskip

\noindent \textbf{Proof of Corollary \ref{cor-CLTlimit}}: The preceding
Lemma proves the extension for Proposition \ref{propn-CLT}.

To prove (\ref{Vv}), define 
\begin{equation*}
v(t,x,y)=\sup_{a\in \lbrack \mathcal{A}](t,1)}E_{P}\left[ u\left(
x+\int_{t}^{1}a_{s}^{(1)}ds,y+\int_{t}^{1}a_{s}^{(2)}dB_{s}^{(2)}\right) %
\right] ,\ \ (x,y)\in \mathbb{R}^{2}.
\end{equation*}%
As in the proof of Lemma \ref{lemma-ddp-G}(1), for $u\in C_{b,Lip}(\mathbb{R}%
^{2})$, it can be checked that {(Yong and Zhou (1999, Theorem 5.2 in Chapter
4))} $v$ is the unique viscosity solution of the HJB-equation (\ref{pde-1})
with function $G$ given in (\ref{G-function}). Then we have 
\begin{equation*}
V=\sup_{a\in \lbrack \mathcal{A}](0,1)}E_{P}\left[ u\left(
\int_{t}^{1}a_{s}^{(1)}ds,\int_{t}^{1}a_{s}^{(2)}dB_{s}^{(2)}\right) \right]
=v(0,0,0)\text{. }
\end{equation*}%
For $u\in C(\mathbb{R}^{2})$ with growth condition, the value function is
still the unique viscosity solution of the PDE (\ref{pde-1}) with function $%
G $ given in (\ref{G-function}). Supporting details can be found in Pham
(2009, p.66) or Aivaliotis and Palczewski (2010, Corollary 4.7). \hfill $%
\blacksquare $

\bigskip \bigskip

\noindent {\Large The Krylov norm}: We{\ use the notation in Krylov (1987,
Section 1.1); see also Peng (2019, Chapter 2.1). For }$\Gamma \subset ${$%
[0,\infty )\times \mathbb{R}^{2}$, $C(\Gamma )$ denotes the set of all
real-valued functions $v$ defined on $\Gamma $, continuous in the relative
topology on $\Gamma $ and having a finite norm, 
\begin{equation*}
\Vert v\Vert _{C(\Gamma )}=\sup_{(t,z)\in \Gamma }|v(t,z)|.
\end{equation*}%
Similarly, given $\alpha ,\beta \in (0,1)$, 
\begin{equation*}
\Vert v\Vert _{C^{\alpha ,\beta }(\Gamma )}=\Vert v\Vert _{C(\Gamma
)}+\sup_{(t,z),(t^{\prime },z^{\prime })\in \Gamma ,(t,z)\neq (t^{\prime
},z^{\prime })}\frac{|v(t,z)-v(t^{\prime },z^{\prime })|}{|t-t^{\prime
}|^{\alpha }+|z-z^{\prime }|^{\beta }}
\end{equation*}%
\begin{equation*}
\Vert v\Vert _{C^{1+\alpha ,1+\beta }(\Gamma )}=\Vert v\Vert _{C^{\alpha
,\beta }(\Gamma )}+\Vert \partial _{t}v\Vert _{C^{\alpha ,\beta }(\Gamma
)}+\sum_{i=1}^{2}\Vert \partial _{z_{i}}v\Vert _{C^{\alpha ,\beta }(\Gamma
)}.
\end{equation*}%
\begin{equation*}
\Vert v\Vert _{C^{1+\alpha ,2+\beta }(\Gamma )}=\Vert v\Vert _{C^{1+\alpha
,1+\beta }(\Gamma )}+\sum_{i,j=1}^{2}\Vert \partial _{z_{i}z_{j}}^{2}v\Vert
_{C^{\alpha ,\beta }(\Gamma )}.
\end{equation*}%
The corresponding subspaces of $C(\Gamma )$ in which the correspondent
derivatives exist and the above norms are finite are denoted respectively by 
\begin{equation*}
C^{1+\alpha ,1+\beta }(\Gamma )\text{ and }C^{1+\alpha ,2+\beta }(\Gamma ).
\end{equation*}%
Therefore, the first and second derivatives $v(t,z)$ with respect to $z$
exist and the related norms are finite. In particular, }$\exists ${$L>0$
such that 
\begin{equation*}
\sup_{(t,z),(t,z^{\prime })\in \Gamma ,z\neq z^{\prime }}\frac{%
|v(t,z)-v(t,z^{\prime })|}{|z-z^{\prime }|^{\beta }}<L.
\end{equation*}%
In the proof of Lemma \ref{lemma-ddp-G}, we applied the preceding to $%
v(t,z)=H_{t}(z)$. }

\bigskip

\noindent \textbf{Completion of the proof of Theorem \ref{thm-tradeoff}(iii)}%
: Show that specializing in arm 2 is not asymptotically optimal if $\alpha <%
\overline{\alpha }$.

Verify the inequality 
\begin{equation*}
\mu _{1}\frac{\sigma _{3}}{\sigma _{1}+\sigma _{3}}+\mu _{3}\frac{\sigma _{1}%
}{\sigma _{1}+\sigma _{3}}-\alpha \frac{\sigma _{1}\sigma _{3}^{2}}{\sigma
_{1}+\sigma _{3}}>\mu _{2}-\alpha \frac{\sigma _{2}^{2}}{2}=E_{P}\left[
u\left( \mu _{2},\sigma _{2}B_{1}^{(2)}\right) \right] \text{,}
\end{equation*}%
and deduce that 
\begin{equation*}
\alpha <\frac{2(\mu _{1}-\mu _{2})\Big[(1-\lambda )\sigma _{1}+\sqrt{%
(1-\lambda )\sigma _{1}^{2}+\lambda \sigma _{2}^{2}}\Big]}{2(1-\lambda
)\sigma _{1}^{3}+(2\lambda -1)\sigma _{1}\sigma _{2}^{2}-\sigma _{2}^{2}%
\sqrt{(1-\lambda )\sigma _{1}^{2}+\lambda \sigma _{2}^{2}}}\equiv 2(\mu
_{1}-\mu _{2})g(\lambda ).
\end{equation*}%
It can be verified that, $g^{\prime }(\lambda )>0$ for $\lambda \in (0,1)$
and $\lim_{\lambda \rightarrow 1}g(\lambda )=\frac{1}{\sigma _{2}(\sigma
_{1}-\sigma _{2})}.$

Therefore, for any $\alpha <\overline{\alpha }=\frac{2(\mu _{1}-\mu _{2})}{%
\sigma _{2}(\sigma _{1}-\sigma _{2})}$, there exists $\lambda _{1}\in (0,1)$
such that 
\begin{equation*}
\alpha <2(\mu _{1}-\mu _{2})g(\lambda _{1})<\frac{2(\mu _{1}-\mu _{2})}{%
\sigma _{2}(\sigma _{1}-\sigma _{2})}.
\end{equation*}%
Choose $\lambda =\lambda _{1}$ in the definition (\ref{ahat}) of $\hat{a}=(%
\hat{a}_{s}^{(1)},\hat{a}_{s}^{(2)})$ and deduce that 
\begin{align*}
V=& \sup_{a\in \lbrack \mathcal{A}](0,1)}E_{P}\left[ u\left(
\int_{0}^{1}a_{s}^{(1)}ds,\int_{0}^{1}a_{s}^{(2)}dB_{s}^{(2)}\right) \right]
\\
\geq & E_{P}\left[ u\left( \int_{0}^{1}\hat{a}_{s}^{(1)}ds,\int_{0}^{1}\hat{a%
}_{s}^{(2)}dB_{s}^{(2)}\right) \right] \\
>& E_{P}\left[ u\left( \mu _{2},\sigma _{2}B_{1}^{(2)}\right) \right] .
\end{align*}%
Therfore, specializing in arm 2 is NOT asymptotically optimal.

When $\sigma _{2}=0$, we can set $\overline{\alpha }=\infty $, and the above
proof still holds.\hfill $\blacksquare $

\end{document}